\documentclass[preprint, authoryear, 5p, number]{elsarticle}
\usepackage{hyperref}
\usepackage{units}
\usepackage{amssymb}
\usepackage[english]{babel}
\usepackage{color}
\usepackage{array}
\usepackage{booktabs}
\usepackage{graphicx}
\usepackage{array}
\usepackage{subfig}
\usepackage{fnpos}
\usepackage{listings}
\usepackage{booktabs}
\usepackage{bbding}
\usepackage{pifont}
\usepackage[dvipsnames]{xcolor}

\usepackage[T1]{fontenc}
\hypersetup{
    colorlinks=true,
    linkcolor=blue,
    filecolor=magenta,      
    urlcolor=blue,
}

\newcolumntype{M}[1]{>{\centering\arraybackslash}m{#1}}
\newcolumntype{N}{@{}m{0pt}@{}}

\newcommand{\HI}{H\textsc{i}}

\newcommand{\kms}{km\,s$^{-1}$}

\newcommand{\xmark}{\ding{55}}%

\journal{Astronomy and Computing}

\begin{document}
	
\renewcommand{\topfraction}{0.95}
\renewcommand{\bottomfraction}{0.95}
\renewcommand{\textfraction}{0.05}

\begin{frontmatter}

\title{FRELLED Reloaded: Multiple techniques for astronomical data visualisation in Blender}

\author[a]{R. Taylor\corref{cor1}}
\ead{rhysyt@gmail.com}
\cortext[cor1]{Corresponding author}
\address[a]{Astronomical Institute of the Czech Academy of Sciences, Prague}

\begin{abstract}
I present version 5.0 of FRELLED, the FITS Realtime Explorer of Low Latency in Every Dimension. This is a 3D data visualisation package for the popular Blender art software, designed to allow inspection of astronomical volumetric data sets (primarily, but not exclusively, radio wavelength data cubes) in real time using a variety of visualisation techniques. The suite of Python scripts that comprise FRELLED have been almost completely recoded and many new ones added, bringing FRELLED's operating environment from Blender version 2.49 to 2.79. Principle new features include: an enormously simplified installation procedure, a more modular graphical appearance that takes advantage of Blender 2.79's improved interface, much faster loading of FITS data, support for larger data sets, options to show the data as height maps in 2D mode or isosurfaces in 3D mode, utilisation of standard \textit{astropy} and other Python modules to support a greater range of FITS files (with a particular emphasis on higher-frequency radio data such as from ALMA, the Atacama Large Millimetre Array), and the capability of exporting the data to Blender 2.9+ which supports stereoscopic 3D displays in virtual reality headsets. In addition, in-built help files are accessible from each menu panel, as well as direct links to a complete wiki and set of video tutorials. Finally, the code itself is much more modular, allowing easier maintainability and, over the longer term, a far easier prospect of migrating to more recent versions of Blender.   
\end{abstract}


\end{frontmatter}

\section{Introduction}
\label{sec:intro}
In \cite{T15} (hereafter T15), I described an earlier version of FRELLED, the FITS Realtime Explorer of Low Latency in Every Dimension. In brief, the FRELLED package takes the form of a set of Python scripts which are run inside the Blender art program (\url{https://www.blender.org/}), providing Blender with the capabilities for loading, examining, and analysing standard radio astronomy 3D data sets. Being designed for artists, Blender provides a very powerful set of tools for data inspection and manipulation, and I showed how these could be utilised to the advantage of astronomers - in particular to facilitate rapid visual cataloguing of reasonably large numbers of sources (i.e. allowing a single astronomer to visually catalogue about 300 sources per working day, as demonstrated in \citealt{agesvii}).

In this paper I describe the capabilities of the latest version, FRELLED 5.0. This has been recoded almost entirely from scratch (the first such time this was done for FRELLED). Only a complete rewrite was feasible as when Blender switched from the 2.4 series (which FRELLED was originally written for) to the 2.5+ versions, its internal Python syntax was completely changed. However, there are many reasons to undertake such a project. Not only is Blender 2.49 effectively obsolete at this point, with no prospect of fixing any of its idiosyncrasies or bugs (let alone implementing new features), but Blender 2.5+ allows access to external Python modules unavailable to earlier versions. This not only greatly simplifies the recoding process by using existing, dedicated modules rather than rewriting standard functions, but also allows the capabilities of the software to be significantly expanded and improved.

The primary aim of this new version of FRELLED remains the same: to allow users to easily view and catalogue their data in 3D. As data sets become ever larger, it is inevitable that some amount of automatic algorithms will have to be employed (e.g. \textsc{polyfind} in \citealt{Polyfind}, GLADoS in \citealt{GLADoS}, \textsc{duchamp} in \citealt{duchamp}, SoFiA in \citealt{SOFIA}; these are all primarily aimed at radio astronomy data cubes as FRELLED is) in order to distinguish signals from noise. But this does not preclude the need for human inspection\footnote{I here limit the discussion to \textit{visual} inspection; for sonification, see e.g. \citealt{sound1}, \citealt{sound2}.}. Larger data sets bring with them the prospect of new, hitherto unexpected discoveries, which will undoubtedly cause difficulties for any pre-programmed algorithm to characterise (but see \citealt{unexpected}). Additionally, when it comes to human visual recognition, it is not the speed at which the brain processes the information (tens to the low hundreds of milliseconds - \citealt{visspeed1}, \citealt{visspeed2}) that forms the bottleneck to cataloguing speed, but how fast the human is able to physically record their discovery: that is, convert the information from visual identification to a catalogue-format entry (typically seconds per source). Thus if the recording speed can be minimised, visual source extraction may yet play a role in cataloguing even the largest of data sets. The quantitative efficacy of visual extraction will be described in detail in a forthcoming paper based on FRELLED (Taylor, submitted), but see also e.g. T15, \cite{AGESVC2}, and \cite{wang}. 

In this work I concentrate on the new capabilities of FRELLED 5.0. Besides new data display methods, particular attention has been given to allowing FRELLED to work with higher frequency data sets such as those from ALMA (the Atacama Large Millimetre Array), as well as greatly simplifying the installation process and providing a significantly more intuitive graphical user interface (GUI): features which previously were implemented but hidden are now included explicitly in their own panels (especially multi-volume rendering). There is a strong emphasis on allowing the user to do as much as possible without touching the code, including procedures which are normally only of use when debugging.

The design philosophies behind FRELLED are intended to tackle an awkward question: why are such tools not more widely used in astronomy ? This is not an issue specific to FRELLED or even Blender in general. Numerous papers describe particular use cases for Blender, see for example \cite{astroblend}, \cite{blendervox}, \cite{noncart}, as well as more general applications such as in \cite{kentpaper, kentbook, kentproc}. Non-Blender visualisation engines have also seen some utilisation, for example, Houdini in \cite{houdini} and Unity in \cite{unityvr, unity}. Why then, with many options available to avoid the difficult task of developing a data visualisation framework, does 3D visualisation remain a relatively niche concern within observational astronomy ?

I suggest two answers to this question with regards to FRELLED. Firstly there is a need for the absolute maximum of \textit{simplicity} in the entire user experience, both in terms of general use but also installation and initial learning. It is not much good providing raw capabilities which are so tedious to understand (or time-consuming to set up) that they are not actually used. This need for simplicity has been termed `fastronomy', \citealt{fastronomy}, while the website for the `Nightlight' FITS viewer (\citealt{nightlight}) expresses things more bluntly: `The answer to the question, “What’s in this FITS file?” should not be “Let me open Python”' (while I agree with this, it is not the only valid design philosophy - see for example \url{https://napari.org/} for a hybrid code and GUI-based file viewer). FRELLED has therefore been (re)designed to ensure simplicity at all stages: once the absolute basics of the operation have been grasped, the remaining features are intended to be as self-explanatory as possible.

The second answer I suggest is \textit{versatility}, both in terms of the available methods of data visualisation and the data sets which can be loaded. The aim is to provide as many different visualisation methods as possible. A series of channel maps may show the same data as a volumetric render, but the subjective viewing experience of the two is very different. Having multiple options within the same tool is designed to give users as many ways to understand their data as is practical, with different features sometimes more easily discernable by one technique than another. Regarding the use of different data sets, FRELLED was previously hard-coded to assume \HI{} data, a severe limitation which has now been removed. More generally, the motivation to provide multiple visualisation techniques is twofold. Firstly, some of these have been directly motivated by my own research where it became apparent that different methodologies are more suitable to particular scientific use cases than others - these are discussed throughout the text as appropriate. Secondly, the goal is to allow users different options when the optimum technique is difficult to anticipate ahead of time. Ideally, users should be given the capabilities to utilise the techniques they already understand but also the flexibility to experiment and find new solutions they might not have otherwise considered.

Both versatility and simplicity are subject to weak links. It does not matter if the use experience is flawless if the installation process is overly-complicated; likewise a viewer only capable of rendering data in one particular way is unlikely to find much use beyond specialised cases. The aim with this version of FRELLED is to overcome all of these possible weak links and provide a general-purpose 3D FITS file viewer, albeit still with the primary target of radio astronomy data sets.

The remainder of this work is as follows. In section \ref{sec:display} I describe the primary new visualisation capabilities of FRELLED, while section \ref{sec:anal} covers the new analysis features (note that the distinction between visualisation and analysis is not strict). Section \ref{sec:interface} covers changes to the interface and outlines the new user guide documentation. Section \ref{sec:code} describes changes to the code. Finally section \ref{sec:sum} summarises the current features of the software and the possibilities for the future. 

Throughout this work, I concentrate on features which are new or substantially changed from the previous versions, which are summarised in table \ref{tab:features}. For full technical details of how FRELLED works, and for description of operations without substantial changes, I refer the reader to T15. The machine used for developing and testing FRELLED is a Lenovo Legion 5 model number 5-15ARH05H, with 16 GB DDR4 RAM, a Nvidia GeForce RTX 2060 6 GB GPU, and a AMD Ryzen 5 4600H 3 GHz CPU (6 cores). FRELLED is registered at the Astrophysical Source Code Library at the following address: \url{https://ascl.net/1508.004}. Installation is supported on Windows and Linux.

\begin{table*}[t!]
\begin{center}
\caption[feature]{Executive summary of the major changes between the last FRELLED release (version 4.5) and the new version presented here. A red cross indicates a feature is not present, an orange tick that it was or is partially supported, and a green tick that it is mature (and/or substantially improved compared to the previous version).}
\label{tab:features}
\begin{tabular}{c c c c}\\
\toprule
\multicolumn{1}{c}{\textsc{Visualisation}}\\
Feature & FRELLED 4.5 & FRELLED 5.0 & Changes and essential notes for FRELLED 5.0\\
\toprule
Volumetric rendering & \color{Green}\CheckmarkBold & \color{Green}\CheckmarkBold &Faster loading\\
2D rendering & \color{Green}\CheckmarkBold & \color{Green}\CheckmarkBold & Faster loading\\
Maximum cube size &	500$^{3}$ & 1500$^{3}$ & Can be increased further but not recommended\\
Displacement maps &	\color{red}\xmark & \color{Green}\CheckmarkBold & \\
Renzograms & \color{Green}\CheckmarkBold & \color{Green}\CheckmarkBold & Faster loading, colour bar\\
Isosurfaces	& \color{red}\xmark & \color{Green}\CheckmarkBold & \\
VR support & \color{red}\xmark & \color{orange}\Checkmark & Limited prototype\\
Multi-frequency support & \color{red}\xmark & \color{Green}\CheckmarkBold & \\	 
Multi-volume display & \color{orange}\Checkmark & \color{Green}\CheckmarkBold & Dedicated GUI\\
N-body display & \color{Green}\CheckmarkBold & \color{red}\xmark & \\
Vector display & \color{Green}\CheckmarkBold & \color{red}\xmark & \\

\toprule
\multicolumn{1}{c}{\textsc{Analysis}}\\
\toprule
Region colour control &	\color{red}\xmark & \color{Green}\CheckmarkBold & \\
World coordinates &	\color{orange}\Checkmark & \color{Green}\CheckmarkBold	& More robust support for different FITS headers\\
NED queries & \color{Green}\CheckmarkBold & \color{Green}\CheckmarkBold & \\
SDSS queries & \color{Green}\CheckmarkBold & \color{Green}\CheckmarkBold & Upgraded to DR17 in 5.0\\
MIRIAD FITS task & \color{Green}\CheckmarkBold & \color{Green}\CheckmarkBold & \\
MIRIAD mask task & \color{Green}\CheckmarkBold & \color{Green}\CheckmarkBold & \\
MIRIAD mbspect task & \color{orange}\Checkmark & \color{Green}\CheckmarkBold &	Full control of all parameters in 5.0\\
Summing flux & \color{Green}\CheckmarkBold & \color{Green}\CheckmarkBold & New method used in 5.0\\
Velocity maps & \color{red}\xmark & \color{Green}\CheckmarkBold & \\
Dispersion maps	& \color{red}\xmark & \color{Green}\CheckmarkBold & \\
Peak flux maps & \color{red}\xmark & \color{Green}\CheckmarkBold & \\
Integrated flux maps & \color{Green}\CheckmarkBold & \color{Green}\CheckmarkBold & \\

\toprule
\multicolumn{1}{c}{\textsc{Interface and Display}}\\
\toprule
Clipping control & \color{orange}\Checkmark & \color{Green}\CheckmarkBold & Directly in the main GUI in 5.0\\
Alpha control & \color{orange}\Checkmark & \color{Green}\CheckmarkBold & Interactive slider in 5.0\\
Multiple shading modes & \color{orange}\Checkmark & \color{Green}\CheckmarkBold & Directly in main GUI in 5.0 \\
Figure creation	& \color{red}\xmark & \color{Green}\CheckmarkBold & \\
Animate data values	& \color{red}\xmark & \color{Green}\CheckmarkBold & \\
Batch rendering (e.g. time series) & \color{Green}\CheckmarkBold & \color{Green}\CheckmarkBold & Dedicated script provided\\
Documentation &	\color{Green}\CheckmarkBold & \color{Green}\CheckmarkBold & Greatly expanded for 5.0\\
\bottomrule
\end{tabular}
\end{center}
\end{table*}


\section{Displaying data in FRELLED 5.0}
\label{sec:display}
The core of FRELLED operations is as follows. The user supplies a FITS file which the code converts into a sequence of PNG images, which are loaded as image textures onto simple plane meshes created in Blender (in Blender a mesh is a type of virtual object consisting of a set of vertices which define faces, which can be assigned materials to show colour and shading). In 3D mode these are shown as a series of adjacent meshes, with each mesh displaying one slice of the data. With the transparency of each mesh set according to the data image values, this gives the appearance of a fully-sampled volume, while in 2D mode only one image is shown at a time. In both cases, to ensure the data is visible from all angles, the data must be `sliced' (converted into PNG sequences) in three orthogonal projections. A script running in the background ensures the correct image (or set of images) is visible based on the current viewing angle. User-adjustable \textit{transfer functions} are used to control how the data values are mapped to both colour and transparency.

\begin{figure}[t]
	\begin{center}
		\includegraphics[width=90mm]{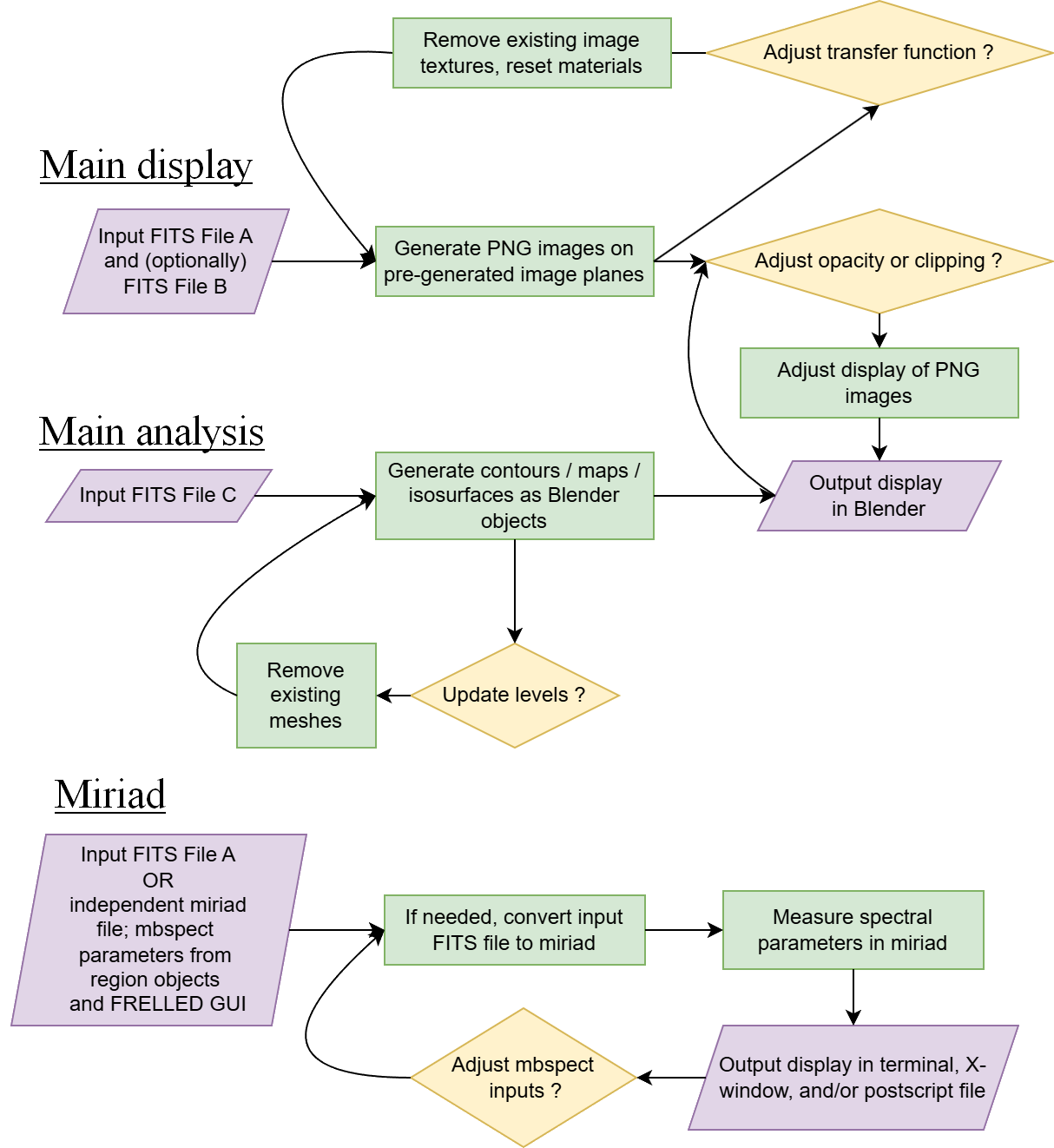}
		\caption[flow]{Simplified flowchart to illustrate how FRELLED displays and analyses data (see section \ref{sec:anal} for analysis operations). Input FITS files are only used in memory (RAM) when accessed, i.e. to convert to PNG or generate contours, whereas the PNG images are retained in GPU memory once loaded.}
		\label{fig:flow}
	\end{center}
\end{figure}

It is important to note that the PNG images are \textit{only} used for visualisation. Whenever FRELLED requires access to the data (such as plotting contours or spectral analysis), it uses \textit{astropy} to access the FITS data directly, rather than using the PNG image files. FRELLED does make occasional adjustments to the FITS data when it imports it for analysis (e.g. ignoring or replacing NaN values) but this is kept strictly internal and the original file itself is treated as read only. FRELLED also retains the scaling convention that one pixel is displayed as equal in size to one Blender Unit\footnote{Later versions of Blender explicitly allow the user to set what one Blender Unit corresponds to in real-word conventions. This is not present in Blender 2.79 but unnecessary for FRELLED; for the choice of using version 2.79, see section \ref{sec:codestruc}.}. A high-level flowchart of the basic operations by which FRELLED displays and analyses data is shown in figure \ref{fig:flow}.

The remainder of this work serves to describe the new features, how they are implemented (and why viable alternative methods were rejected), and their uses for astronomers. Actual guides to using them are given online, see section \ref{sec:readthedocs}. 

\subsection{Volumetric display} 
\label{sec:voldisp}
This occurs as described above, with two important differences from earlier FRELLED versions. First, the conversion from FITS to PNG is done using the Python Image Library (PIL; \citealt{PIL}) rather than \textit{matplotlib} (\citealt{matplotlib}). The reason for this is speed. PIL does not have to set up any axes, image boundaries or anything else superfluous to the image data itself. The result is that each individual slice of the data can be converted to an image about a factor of six faster than using \textit{matplotlib} (this is for the largest data sets as described in section \ref{sec:bench}, for smaller files this can rise to a factor of twelve). As in previous versions, note that each complete PNG sequence is generated using parallelisation with Python's \textit{parmap} module.

\begin{figure*}[t]
	\begin{center}
		\includegraphics[width=180mm]{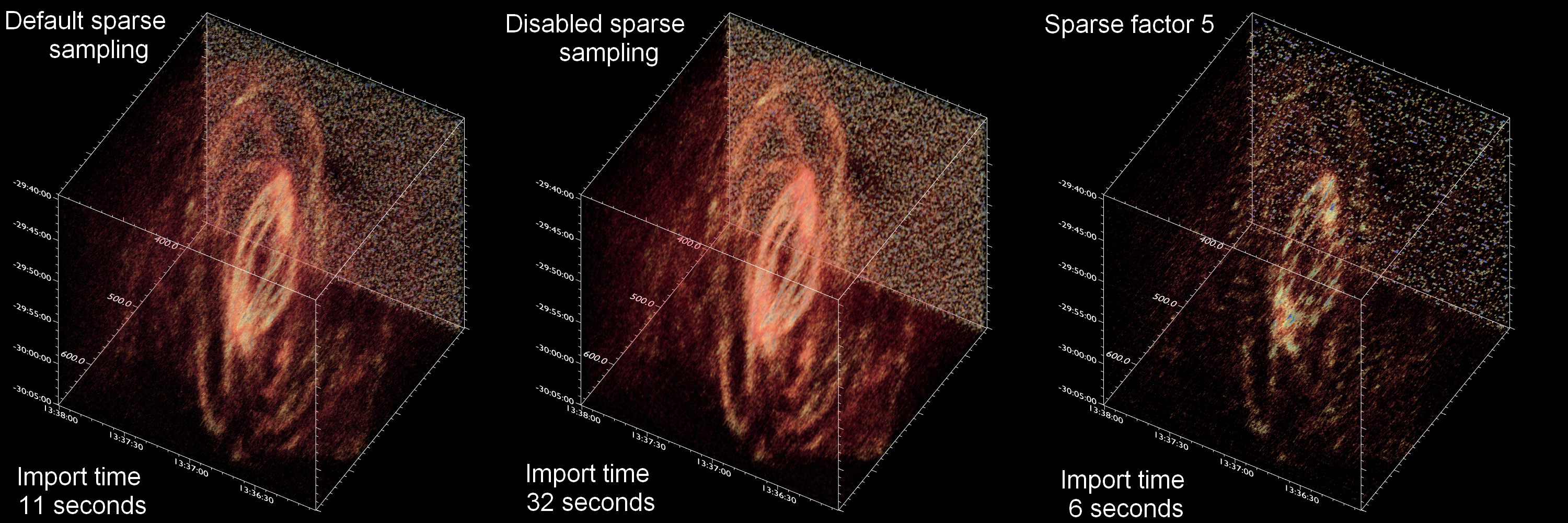}
		\caption[sparse]{Illustration of the different image quality with and without different levels of sparse sampling, using the THINGS \HI{} data cube for NGC 5236 (\citealt{THINGS}). Using either the default (left) or disabled (middle) sparse sampling results in no appreciable difference in image quality, except when zooming in, for a difference in loading times of almost a factor of three. Using a high sparse factor (right) gives faster previews (considerably faster, depending on the data set) but at the expense of quality. The import times quoted are for all three projections of the data, hence the non-linear values of import speed as the sparse sampling does not affect all projections equally; additionally, this neglects the 13 seconds required in each case to load the FITS file into memory and generate the axes.}
		\label{fig:sparse}
	\end{center}
\end{figure*}

The second difference is termed \textit{sparse sampling}. This means that instead of loading every slice of the data in all projections, only every \textit{n}\textsuperscript{th} slice need be loaded (striding across the missing channels), where the value of \textit{n} is allowed to vary between different projections. This is a significant optimisation with the speed benefits being obvious. One might expect that omitting large fractions of the data would have a detrimental effect on the apparent image quality, but as shown in figure \ref{fig:sparse}, in practice this is not necessarily so. While the user can manually increase \textit{n} to allow faster previews of their chosen transfer functions, which can indeed come at the expense of image quality, the default value of \textit{n} is chosen more carefully, as explained below. Crucially, however, sparse sampling does necessarily omit data, and if users require maximum accuracy with every pixel to be shown, it can be disabled.

Key to this is understanding how both Blender and FRELLED determine opacity. Consider one single image slice. In Blender, this will have an opacity value which is set as a property of its material. This material transparency is a uniform value; if the material is a simple single colour value, then the image plane will have exactly this opacity value everywhere. However in FRELLED the material also has a \textit{texture}, which is the image file corresponding to a slice of the data. The values of this image texture are used to modify (by default multiply) the opacity values at all points throughout the plane. The multiplication is by default restricted to the range [0.0..1.0]. A material with an opacity of 0.1 would therefore never be able to have an opacity exceeding this, but may well have parts where the opacity is actually 0.0. Many radio astronomy data sets are dominated by noise, where a well-chosen transfer function will indeed display them as being completely transparent.

FRELLED operates on the simplifying assumption that the opacity of every slice always corresponds to its maximum possible value, the material opacity property. This can be set within the the FRELLED GUI with an adjustable opacity slider, which accounts for the number of images slices being displayed simply as:
\begin{lstlisting}
	material.alpha = int(GUI.alpha/n_slices)
\end{lstlisting}
Where \textit{material.alpha} is the actual opacity value of the material, \textit{GUI.alpha} is the value the user controls in the GUI, and \textit{n\_slices} is the number of images slices. This scheme ensures the actual, on-screen opacity for most of the data is typically low, while only the bright regions (i.e. sources the user is interested in) are clearly visible but without obstructing the view of the data behind them along the line of sight. It gives the user control over the display of the whole data using a single value.

However, observational data sets are rarely perfect cubes. There is no constraint on observers to survey perfectly square patches of the sky, let alone any insistence that the spectral axis have any correspondence with the angular dimensions of a survey. This means that the value of \textit{n\_slices} varies from projection to projection. Even if the value of \textit{material.alpha} is adjusted to account for this, as one rotates the view in FRELLED, there is a very noticeable change when changing from one projection to another. This occurs because the greater the difference in \textit{n\_slices}, the greater the difference in \textit{material.alpha}: individual parts of the data are therefore rendered on-screen with very different opacity values. For example in the THINGS (\citealt{THINGS}) data sets, the spatial resolution may give, say, typically 1,000 pixels on a side, but there might be only 50 spectral channels, thus giving \textit{material.alpha} values which differ by a factor of 20.

By attempting to equalise the number of image slices along all projections, sparse sampling minimises this effect and allows a single value of \textit{GUI.alpha} to correspond reasonably accurately with \textit{material.alpha}, while of course greatly improving the loading times. The method used is to try to equalise the number of images to that of the shortest projection as follows:
\\
\begin{lstlisting}
	ortho_axis_step = int(n_axis/smallest_axis_size)
\end{lstlisting}
Where \textit{n\_axis} is the length of any particular axis, \textit{smallest\_axis\_size} is the length of the shortest axis, and \textit{ortho\_axis\_step} is the step between each rendered image along the axis perpendicular to that under consideration. For example, FRELLED comes with an AGES \HI{} data cube of M33 (Arecibo Galaxy Environment Survey; see \citealt{ages} for the whole survey and \citealt{AGESM33} for the specific data set) as a demonstration file for new users. This has dimensions of 300$\times$270$\times$80 pixels. Thus the shortest axis of this cube is the spectral axis, which has 80 channels. With the default sparse sampling, for this projection all 80 slices of the data would be shown, whereas \textit{ortho\_axis\_step} would be 3 for both other projections. This corresponds to producing and displaying 90 and 100 data slices in each of these projections.

In addition, the user can also alter the sparse factor manually (as well as disable it completely if required), allowing for quicker previews which can be especially valuable when loading larger data sets. This is done by multiplying all values (that is, for all projections) of the \textit{ortho\_axis\_step} parameter by a number the user can enter in the GUI. Altering the sparse factor requires the data to be reloaded.

The combination of replacing \textit{matplotlib} with PIL and sparse sampling means the typical loading speed of a FITS file in the new FRELLED is always faster than in previous versions. The extent is highly variable, typically by a factor of a few, but can exceed an order of magnitude in some cases. There are some minor caveats to this, with small data cubes not loading much faster than previous versions, but with some important compensating factors - these are discussed further in section \ref{sec:bench}.

\subsection{Channel maps} 
\label{sec:chanmaps}
As well as its volumetric display, FRELLED can also show 3D files as a sequence of 2D images along each projection (i.e. channel maps and position-velocity diagrams). Loading data in this way has been optimised from previous versions. In earlier versions, each map was loaded on a separate mesh object. The display of these was animated so that the appropriate slice would be visible based on the current frame in Blender's timeline. This operation has been greatly simplified: a single mesh object is now used with an animated texture, which updates the image shown as the user changes frame. Again the advantage is speed and more efficient memory use. In 2D display mode there are now only three mesh objects required, one per projection, regardless of the size of the data. Limitations on the display capabilities are therefore strongly system-dependent, whereas for earlier versions the bottleneck was Blender itself, as Blender's performance degrades non-linearly with the number of objects (see section \ref{sec:sizematters}).

\subsection{Basics of data inspection}
FRELLED was developed as part of AGES. The combination of volumetric and channel map displays have formed the mainstay of visual source extraction techniques for AGES and its successor WAVES (Widefield Arecibo Virgo Environment Survey; \citealt{waves}) since its initial development. Our standard technique is to first inspect the data volumetrically, mask identified sources using \textit{regions} (see section \ref{sec:regions}, but at their simplest, these are cuboid objects which mark the position of an identified feature), and then search the masked data but now shown as channel maps.
	
This dual approach  balances the advantages and disadvantages of each visualisation method. The volumetric display shows the 3D structure of the sources in their entirety, which makes defining a mask region relatively straightforward and fast. The main disadvantage is that the volumetric display in Blender can only show the sum of the values along the line of sight, meaning that weaker sources tend to be obscured by the noise. In contrast individual data slices (e.g. channel maps) can show fainter data values without obscuration, but accurately defining masks can be more laborious as the user cannot see the full extent of the source in each map: they must pan through the data to find the maximum limits of the source, and also to adjust the position of the region to ensure it does not mask more of the data than is necessary. Using these two visualisation methods in combination ensures we can rapidly mask the brighter sources while taking longer for the fainter ones which require a more careful examination.

\subsection{Isosurfaces and renzograms} 
\label{sec:iso}
Earlier versions of FRELLED had the capability of displaying renzograms (\citealt{renzo}), which plot contours at a single level over multiple channels. The present version retains this capacity, but also allows for true isosurfaces: surfaces plotted at a constant data value. While these by definition rely on clipping large amounts of data, whereas volumetric displays do not, in practise this can have significant benefits. In \cite{fading} we showed that using renzograms can be a powerful way to detect small extensions in marginally resolved galaxies, which would otherwise be hidden by a combination of the brightness of the main source and the obscuration - in a volumetric render - from the surrounding noise. By restricting the view to only flux of a given significance level, finding such features becomes much easier. Indeed some viewers rely exclusively on isosurfaces (\citealt{slicer}), which also possess the significant advantage of being straightforward to export into other 3D object formats.

Isosurfaces are implemented as a type of map in FRELLED (see section \ref{sec:regions}) but can also be shown for the entire data set. How complex they can be is system-dependent: on the development machine, objects can be generated of a few million vertices in about a minute. FRELLED uses the \textit{sckit-image} Python module (\citealt{scikit-image}) using the Lewiner `marching cubes' algorithm (\citealt{marchcubes}) for isosurface generation, with additional code to remove points returned at NaN positions which would otherwise cause problems when trying to load the vertices into Blender. 

When displaying data volumetrically, speed is dominated by the dimensions of the data cube, with only a weak dependence on the distribution of data values (see section \ref{sec:bench} for details). The speed of isosurface generation, in contrast, is strongly dependent on the properties of the data set as well as size. A large cube with a surface generated for only, say, the brightest 1\% of data values will load much more quickly than a surface enclosing 50\% of the values, since the former requires far fewer polygons to be generated. The highly individual, arbitrary nature of the data distribution in any cube means that a generalised quantification is impossible, but in some circumstances, large cubes can be faster to load as isosurfaces than as volumetrics.

\begin{figure*}[t]
	\begin{center}
		\includegraphics[width=180mm]{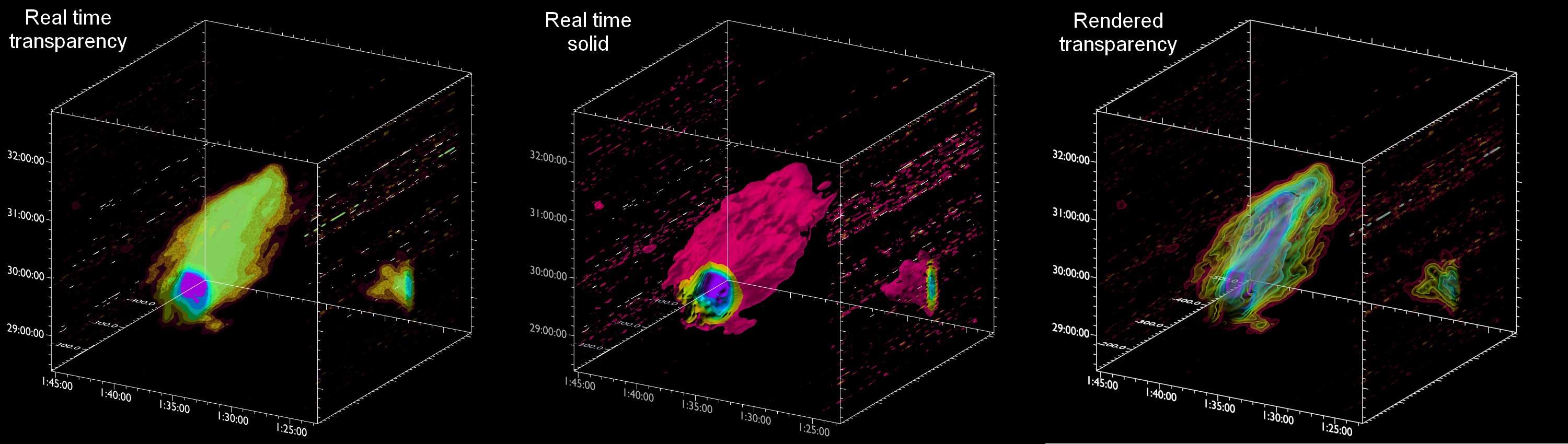}
		\caption[iso]{Isosurfaces for the AGES M33 \HI{} data cube included with FRELLED, shown using different render modes. The left panel shows the default real time view where the surfaces are shown with transparency but with noticeable artifacts. These can be avoided either by deactivating transparency (middle panel) or by rendering the figure using Blender's internal rendering engine (right panel).}
		\label{fig:iso}
	\end{center}
\end{figure*}

A significant limitation for isosurfaces is that Blender's transparency in the real time display (here meaning any operation where the display is updated automatically) shows significant artifacts for complex objects. Several options are available to overcome this. In the rendered view there is no difficulty, and Blender 2.79 allows this viewing mode as a preview in the real time display, albeit at the expense of the display speed. Alternatively the viewing mode can be switched to a simpler view which shows the surfaces without transparency, but retaining the colours. The user can use the alpha clipping control (see section \ref{sec:opacity}) to constrain which surfaces are visible, so all the isosurfaces can be shown separately. Examples of isosurfaces using the different display methods are shown in figure \ref{fig:iso}; for reference, the total vertex count here is about 200,000.

Whether one should use renzograms or isosurfaces for data inspection depends on the scientific use case. For the small tails described in \cite{fading}, renzograms were more appropriate: they show at a glance how many channels a potential source is present in, giving direct information on their statistical significance. For larger, more complex sources, such as the clouds around M33 (some of which are shown in figure \ref{fig:iso}), renzograms are less helpful. In that case the complex structure and large number of channels in each source quickly becomes visually confusing, whereas the simpler appearance of an isosurface is much easier to intuitively understand.

\subsection{Height maps} 
\label{sec:height}
In 2D mode, by default the maps are shown as ordinary images. A new feature allows the display to be toggled to height maps (also known as height fields or displacement maps), in which each pixel is displaced, orthogonally to the plane, by an amount depending on its data value. While Blender does include a native ability to displace meshes according to image texture values, this is not used here - for the sake of accuracy, the FRELLED scripts access the original FITS data and use the actual data values for the displacement, rather than the corresponding image texture pixel values.

Height maps can be shown with or without their accompanying image textures. The amount of displacement is here somewhat arbitrary as flux units can of course be extremely small or large numbers that may not scale well in relation to the pixel units Blender uses for scale. A numerical slider allows the user to adjust this scaling interactively, and a drop-down menu allows different scaling functions to be used (linear, logarithmic, or square root). This ability to change the mesh scaling means that height maps offer an important advantage over standard image textures, as Blender does not allow the transfer function of image textures to be changed after loading, whereas height maps can be rescaled freely.

\begin{figure*}[t]
	\begin{center}
		\includegraphics[width=180mm]{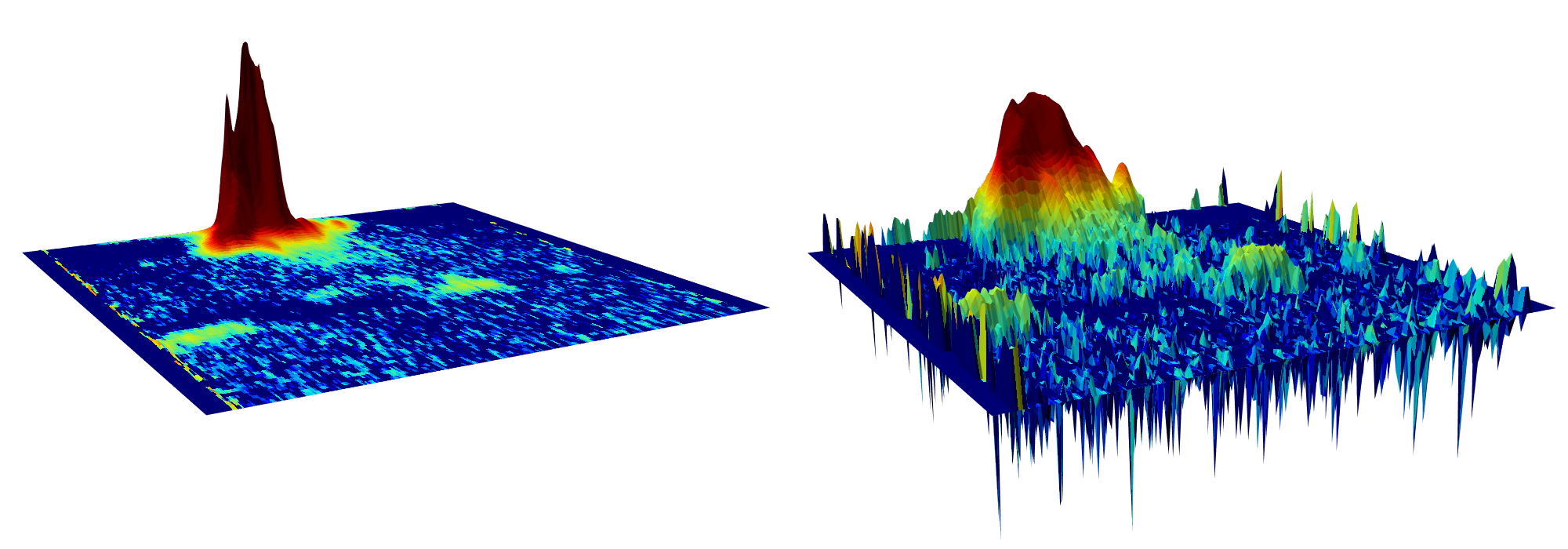}
		\caption[height]{Height maps of a position-velocity slice of the supplied M33 data cube using arbitrary scaling. The left panel uses the default linear scaling while the right panel using logarithmic scaling for the displacement.}
		\label{fig:heights}
	\end{center}
\end{figure*}

Another related advantage of height maps is that they do not suffer the same limitations of displaying data of high dynamic range as with standard images. With highly variable data, it is inherently difficult to generate an image that shows both the brightest and faintest features simultaneously. Typically, either the faintest features are clearly visible but the brightest are saturated, or the brightest features are clear but the faintest are `flattened' into homogenity. Notably, this is the case for the M33 data cube supplied with FRELLED. Consider figure \ref{fig:heights}, which shows M33 with height maps using different transfer functions. In particular in the left panel, the faintest clouds (blue-green) have barely any visible displacement, whereas the bright emission from M33 itself is a uniform dark red but with clear structure in its displacement.

Using displacement also allows the user to interactively explore features of different dynamic range: one can simply move the viewpoint to show the features at the flux scale one is interested in. Of course there are limitations of convenience depending on how extreme the data differences are (moving to a peak a million times further away than the faintest would be awkward~!), but fortunately most data does not show such dramatically different value scales. Note also that the transfer function used for generating the image texture and for the displacement are independent, so one could, for example, use a logarithmic image texture but a linear displacement function. Other ways to address high dynamic range issues are discussed in section \ref{sec:multi}.

\subsection{Virtual reality} 
\label{sec:vr}
The increased popularity of stereoscopic headsets has not gone unnoticed by the astronomical community. For example, \cite{VR1}, \cite{VR2}, \cite{VR3} and \cite{VR4} describe particular cases of viewing astronomical data using different commercially-available headsets and a variety of software packages. However as noted in those publications, especially \cite{VR2}, there is a lack of dedicated astronomer-friendly VR software, and VR use remains by no means widespread. An important recent development in this regard is the iDaVIE software (\citealt{idavie1}, \citealt{idavie2}) an astronomical viewing and analysis tool dedicated to displaying data in virtual reality. Blender 2.79 does not have the ability to directly display its real time view in VR, but FRELLED comes supplied with two scripts to allow export of the data to later versions of Blender which do.

There are several motivations for displaying data in true 3D besides the underlying drive in FRELLED to offer as many viewing techniques as possible. The standard 3D view of the data allows projection effects, where structures are detected as false positives due to chance alignments, to be greatly reduced compared to traditional 2D displays of individual data slices: in 3D one can simply rotate the data slightly and such features disappear. Having a true stereoscopic 3D view means that one can see depth directly, reducing alignment effects still further. This also makes it possible to precisely locate features in 3D space more easily. Of course, this more natural way of viewing the data has an obvious appeal for outreach: one can step inside a data cube, or view it as a scalable object, without needing any dedicated physical objects to represent the data (but for 3D-printed tactile data products, see \citealt{tactile}).

In FRELLED this feature is best regarded as a proof-of-concept or prototype than a fully-developed feature. Almost uniquely, it requires the user to edit the text of the code rather than interacting with the GUI. However, this is sufficient to examine the performance and use cases for Blender as an astronomical VR tool. Displaying the data in VR is a two-stage process: first the user runs one of two scripts inside FRELLED that processes the file, writing a list of all the objects (either isosurfaces or image planes for volumetrics) that need to be exported to a text file along with some material parameter values. A second script must then be run in the special supplied Blender file designed for Blender 2.91, which loads the objects and converts the materials into a format suitable for the real time engine used in 2.91 (this also applies to all subsequent Blender versions, see section \ref{sec:sum}). Then the user can simply start a VR session inside Blender using the built-in `Scene Inspector' add-on. 

As a prototype, the scripts simply export and import the data, and no corresponding background script has been written for 2.91 that adjusts which set of meshes should be visible based on the viewing angle (see section \ref{sec:display}; it is unclear if such a script would be able to account for the headset position rather than the ordinary user viewpoint). For isosurfaces this makes no difference, but for volumetric data, the user must either accept a limited range of viewing angles, or import all the volumetric meshes onto the same layer. This results in a reduction in frame rate.

The experience of using Blender in this way might be described as only sufficient for small volumetric cubes, but more than adequate for isosurfaces. It should be stressed that it does not compare with the dedicated engine of iDaVIE, which loads the data very much faster (see section \ref{sec:bench}). Nevertheless, subjectively the stereoscopic capability does appear to add something significant to the viewing appearance, making features more easily visible than in the standard flatscreen view. Cautiously, it does seem that working fully within VR has potential. While Blender's engine is of lower performance than iDaVIE, it possesses the advantage of being able to view data in multiple modes (simultaneously if need be), e.g. as transparent isosurfaces, solid annotations (see section \ref{sec:figs}), or the textured planes used for volumetrics. For outreach, one can see the height fields as solid textured surfaces, enabling the user to walk around a virtual landscape composed from a FITS file (height fields do not yet have a dedicated VR export feature as they can simply be appended as a single mesh object).

\begin{figure}[t]
	\begin{center}
		\includegraphics[width=90mm]{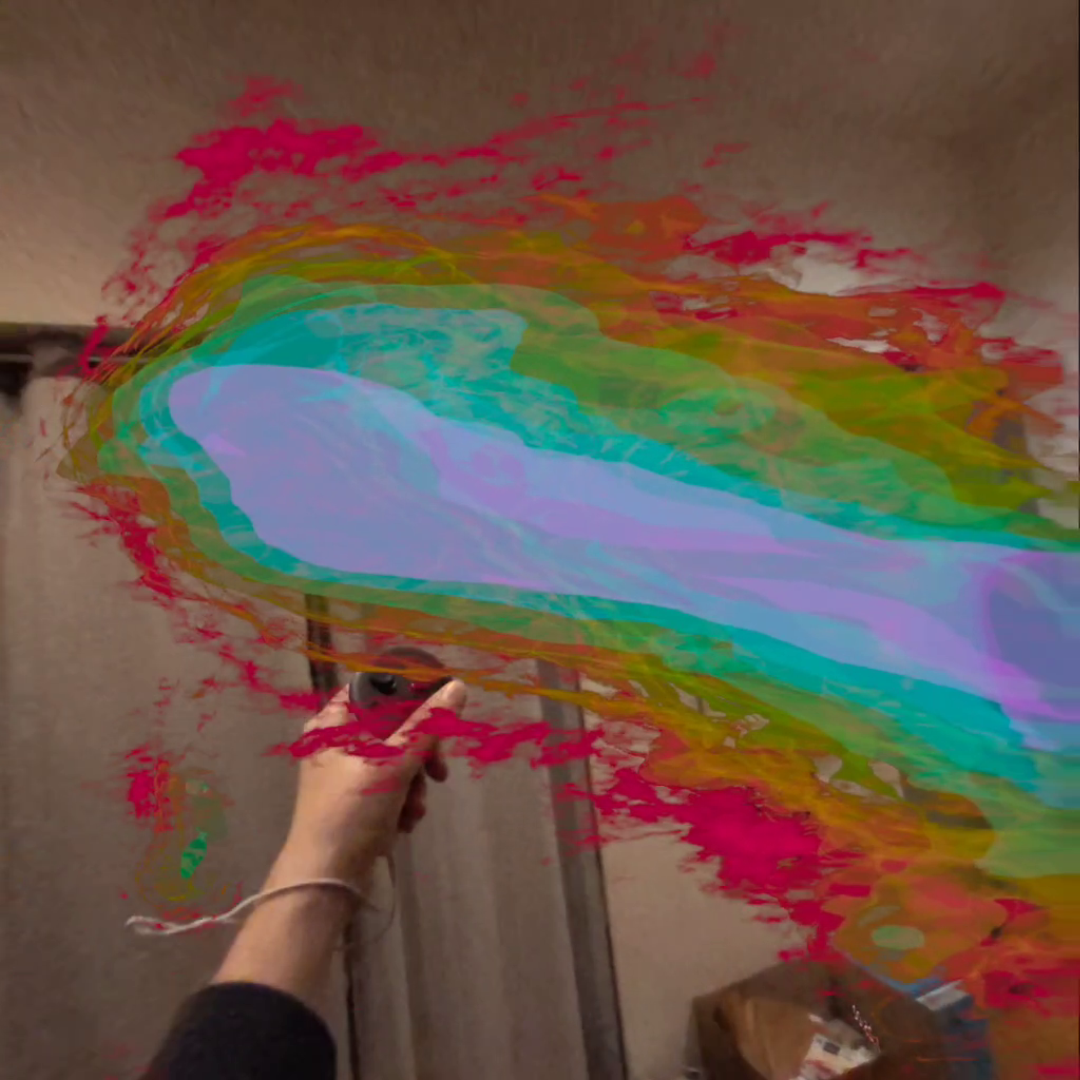}
		\caption[VR]{Example of isosurfaces in mixed reality, captured on the author's Meta Quest 3 headset using the Virtual Desktop software. An accompanying video can be found in the supplementary information. In Blender 2.91 the user can only move around the data set by physically moving, and the hand-held VR controllers do nothing. In Blender 4.0 the controllers, represented as white spheres, can be used to translate and rescale the whole Blender scene (i.e. the data set) interactively.}
		\label{fig:VR}
	\end{center}
\end{figure}

Mixed (or augmented) reality is also possible. An example of this, using isosurfaces generated from the default M33 file, is shown in figure \ref{fig:VR}. This is enabled with the commercial software `Virtual Desktop' (\url{https://www.vrdesktop.net/}). In a standard VR display mode, Blender shows the data against a uniform background colour. Virtual Desktop uses the chroma key method to replace a specified colour with the camera passthrough of the headset (when available). Since this colour can be black (as set by FRELLED by default), it is not necessary to carefully choose the colour scheme for the data, since black is normally used for the zero-opacity data anyway. The virtual image is overlaid on the passthrough image and occlusion by foreground objects not yet supported in Virtual Desktop. Virtual objects do not interact in any way (e.g. lighting) with the real world environment. This makes the experience somewhat crude, but mixed reality possess the significant advantage over virtual reality that new users are continuously aware of the real world and can therefore move around more freely. Given that the Quest natively supports occlusion, and the existence of user experiments which include dynamic lighting effects in the passthrough, it is not unreasonable to expect more engaging experiences in the very near future.

\subsection{Display limitations} 
\label{sec:limits}
\subsubsection{Loss of features} 
In recoding FRELLED, the new features and coding style (see section \ref{sec:code}) approximately doubled the total length of the code and necessitated more extensive debugging. To ensure a timely release and avoid a permanent state of development, prioritisation was given to features most important for FRELLED's main intended purpose of analysing observational data sets. Thus the major compromise has been the loss of the n-body and vector viewing facilities of the older versions. Re-implementing these is anticipated to be a relatively small workload but would depend on user interest. 

\subsubsection{Data cube size restrictions} 
\label{sec:sizematters}
For optimising the speed of displaying volumetric data, FRELLED 5.0 uses the same setup as previous versions, with the file containing pre-loaded blank image planes (see T15 section 2.6). This greatly reduces the time needed to load the data as all the scripts have to do is edit the material texture settings and reposition the planes, which is very much faster than creating objects and materials from scratch. Previous versions of FRELLED came with 500 blank planes per projection. In the new version this has been increased to 1,500\footnote{Each projection requires image plane meshes for both the forward and reverse viewing angles (though these use the same PNG images, see T15). This means that there are actually 3,000 blank planes per projection, so 9,000 in total.}. This reflects more modern computing power as well as the increased demand for viewing larger data sets. The more objects present within a Blender scene, the lower the performance, and this value was found to give an acceptable balance between allowing for large data sets and having a real time display that ran at a high frame rate. Unlike earlier versions, this value can be increased by the user, see section \ref{sec:configvol} (if a user tries to load a file which is too large, by default FRELLED automatically erases the blank planes and creates a new set of the appropriate number; see section \ref{sec:configdel}). 

In short, data sets can be loaded which are up to 1,500 pixels on a side as standard, but it is possible to increase this. Besides actually increasing the number of blank planes there are two options. First, there is no hard-coded limitation at all on the data size which can be loaded in 2D mode. As discussed in section \ref{sec:chanmaps} this mode now uses only one image plane per projection, which uses negligible system resources: the only restriction is how large an image file the user's system can process and display, which is unrelated to the FRELLED code. 

Secondly, the asymmetrical nature of typical data sets, while causing some problems addressed in section \ref{sec:voldisp}, can also be advantageous: as in 2D mode, for the volumetric display there is no restriction on the size of individual images which can be loaded. For example in the case of the highly asymmetrical AGES cube used in section \ref{sec:bench}, while over 15,000 images would be required along the sky projection, only about 300 are needed along each of the position-velocity projections. FRELLED imposes restrictions along each projection separately, which means that for data sets like this, one can at least load individual projections, just not necessarily all of them.

\subsubsection{Colour transfer function limitations} 
As with earlier versions, it is not possible to alter the colour or opacity transfer functions interactively. While the faster loading speed reduces the burden of having to reload data if the scaling was not optimal, this is not an ideal solution; sparse sampling only alleviates the problem by allowing faster loading speeds but by no means eliminates it. I note that in more recent versions of Blender, it is possible to interactively alter the transfer functions, as can be done when exporting the data for VR display in Blender 2.91. There is also a tension between the inherent need for higher total opacities (which results in brighter, more visible data) against the need to avoid obscuring the view, and this cannot be fully resolved without altering Blender's source code for its real time display. A better option would be to display the image as the peak value along the line of sight rather than the sum (or, ideally, utilise the more sophisticated transfer functions described in \citealt{vohl}).

\subsection{Benchmarks and performance comparisons} 
\label{sec:bench}
Some performance benchmarks for different data cubes are presented in table \ref{tab:bench}. This shows the results for both real observational data and random-noise simulated data. Cubes larger than 1,024$^{3}$ voxels were not tested due to memory limitations, and in any case at a real time display of 4 f.p.s. (frames per second) the file becomes cumbersome to work with. Note however that this is non-linear, with smaller cubes being much faster to both load and display. For the hardware and software necessary to deal with even larger data sets, see \cite{reallybigdata}.

Direct comparisons with the earlier versions of FRELLED are difficult due to the change in hardware. However one previous test machine, the HP Elite 7500 series desktop (with an Intel i7-3770 quad core 3.4 GHz CPU, 16 GB RAM and a 4GB Nvidia GeForce GT 640 GPU), is still active. On that machine, cubes of 577$^{3}$ voxels previously took over five minutes to load whereas cubes of 512$^{3}$ voxels now take approximately 45 seconds on the (current) development machine. The same 512$^{3}$ voxel cube on the HP Elite, using FRELLED 5.0, takes a very similar time to load as on the development machine of 47 seconds. This indicates there has indeed been a large overall performance improvement thanks to the optimisation of the code in this latest version of FRELLED.

The relatively poor real time performance of only 15 f.p.s. for the random-noise 512$^{3}$ voxel cube is because the random-noise data is essentially a worse-case scenario for the display. With signal being present along all lines of sight, the GPU is constantly integrating to calculate the image. In real use cases, such a performance can be expected (for example) in Galactic \HI{} cubes in which signal is present in a majority of voxels. The comparable-size AGES and PHANGS data sets are more representative of typical \HI{} and CO extragalactic data in which much of the volume has negligible signal, and accordingly the display performance is significantly better. 

The tests given in table \ref{tab:bench} were carried out using the default FRELLED options including sparse sampling, but two special adjustments had to be made. For the extremely asymmetrical AGES NGC 7448 data set, as mentioned in section \ref{sec:sizematters}, only two projections were loaded due to the limitation of only allowing 1,500 images along each projection. For the simulated data of 1,024$^{3}$ voxels, memory limitations forced a modification of the code to only use three cores for converting the FITS data to PNG files, resulting in a much slower process than usual.

\begin{table*}[t!]
\begin{center}
\caption[bench]{Performance benchmarks for selected data sets. The AGES M33 cube is described in \cite{AGESM33} while the AGES NGC 7448 cube is a 300 MHz bandwidth cube using unpublished data. The PHANGS data is presented in \cite{phangs}. The random-noise data sets were generated simply using the numpy.random.normal function in Python. For the 1,024$^{3}$ voxel cube, memory limitations necessitated restricting the FITS to PNG conversion from the usual 6 to 3 cores, hence the much slower loading time than the other data sets. Here, `loading time' refers to the time from pressing `load data' in FRELLED to having the data visible on-screen. All three projections were loaded for all data sets except for the AGES Mock data, for which only the two position-velocity projections were used. `Reloading time' refers to the time to load the FRELLED file after having already saved it with the converted data. For observational data sets, each row gives the file loaded, its exact dimensions and equivalent cube root, and file size in megabytes.}
\label{tab:bench}
\begin{tabular}{c c c c}\\
\toprule
\multicolumn{1}{c}{\textsc{Observational data sets}}\\
Data & Loading time & Reloading time & Display performance\\
\toprule
AGES - M33 & FRELLED : 12\,s & 3\,s & $>$\,25 f.p.s.\\
330$\times$270$\times$80 & GLNemo2 : $<$\,1\,s & & \\
7,992,000 voxels $\equiv$ 200$^{3}$ & iDaVIE : $<$\,1\,s & & \\
25\,MB & & & \\
\cmidrule{1-4}
PHANGS - NGC 2903 & FRELLED : 28\,s & 8\,s & $>$\,25 f.p.s.\\
816$\times$1,102$\times$221 & GLNemo2 : 38\,s & & \\
198,730,272 voxels $\equiv$ 584$^{3}$ & iDaVIE : 3\,s & & \\
776\,MB & & & \\
\cmidrule{1-4}
AGES - NGC 7448 Mock & FRELLED : 38\,s & 10\,s & $>$\,25 f.p.s.\\
300$\times$100$\times$15,624& GLNemo2 : 3\,m 34\,s & & \\
468,720,000 voxels $\equiv$ 777$^{3}$ & iDaVIE : 5\,s & & \\
1.8\,GB & & & \\
\bottomrule
\\
\toprule
\multicolumn{1}{c}{\textsc{Random-noise data sets}}\\
\toprule
128$^{3}$ voxels & FRELLED : 13\,s & 3\,s & $>$\,25 f.p.s.\\
16.4\,MB  & GLNemo2 : 1\,s & & \\
 & iDaVIE : $<$\,1\,s & & \\
\cmidrule{1-4}
256$^{3}$ voxels & FRELLED : 20\,s & 5\,s & $>$\,25 f.p.s.\\
131.1\,MB  & GLNemo2 : 37\,s & & \\
 & iDaVIE : 2\,s & & \\
\cmidrule{1-4}
512$^{3}$ voxels & FRELLED : 45\,s & 9\,s & 15 f.p.s.\\
1.0\,GB  & GLNemo2 : \textit{Crashed} & & \\
 & iDaVIE : 3\,s & & \\
\cmidrule{1-4}
1,024$^{3}$ voxels & FRELLED : 6\,m 28\,s & 33 s & 4 f.p.s.\\
8\,GB  & GLNemo2 : \textit{Crashed} & & \\
 & iDaVIE : 16\,s & & \\
\bottomrule
\end{tabular}
\end{center}
\end{table*}


Loading times for small data sets, such as the default M33 example file, do not benefit from the same improvements in loading time as for larger files. Indeed they may load no more quickly than in previous versions of FRELLED, due to the larger number of objects present in the default file (see section \ref{sec:sizematters}). The trade-off from not reducing the loading times for the smallest files (by at most a few more seconds) is that the files loaded can now be 27 times larger in volume, which seems to be a very reasonable exchange. 

While loading speeds of $\sim$\,10\,seconds are no great burden in most typical situations, it must be admitted that times on the order of a minute do become inconvenient if the data range used for the transfer function was not initially suitable: one minute can easily become many times longer. FRELLED helps the user by providing estimates of the percentile ranges of the data directly in the GUI, and also printing the \textit{rms} value (at 1, 2, 3 and 4$\sigma$ multiples) to the terminal. But this by no means guarantees a successful data range. Loading speeds in \textit{kvis} for files comparable to the largest considered here are of the order of mere seconds, but this is not a fair comparison as this only shows the data as a series of 2D images. 

Better comparisons can be made with other dedicated 3D viewers, with the loading speeds for both GLNemo2 (\citealt{glnemo}) and iDaVIE also given in table \ref{tab:bench}. Overall, FRELLED improves on the loading speeds of GLNemo2 for all but the smallest cubes, though iDaVIE is very much faster than FRELLED. Unlike the other viewers, FRELLED can save and reload the whole work state, and these reloading speeds are comparable with iDaVIE's loading speeds. It should also be noted that while GLNemo2 and iDaVIE do not allow direct frame rate measurements, it was clear that there was a significant performance drop in both viewers when viewing the Mock data set while FRELLED maintained a high frame rate - the solution here of loading only two projections of the data is not ideal, but is at least a useful option the other viewers do not possess. iDaVIE also struggled with the 1,024$^{3}$ random-noise data set, though was noticeably better than FRELLED. However, in terms of both loading speed and real time frame rate, iDaVIE is a decisive overall winner, especially given that iDaVIE allows for real time control of the transfer function while FRELLED does not.
	
On the other hand, iDaVIE strictly requires the use of a VR headset, whereas both FRELLED and GLNemo2 are primarily aimed at conventional PCs. A broader comparison of the qualitative features offered by these and other visualisation tools is given in section \ref{sec:notbugs}.

Finally, it is difficult to extrapolate how FRELLED's performance would scale with different hardware. Nonetheless, it may still be helpful to compare the performance on a limited set of different machines. The following loading times are for the random-noise 512$^{3}$ voxel cube:
\begin{itemize}
\item Acer Aspire A515-54G with Intel Core i7 1.8 - 2.3 GHZ CPU, 16 GB RAM, integrated Nvidia GeForce MX350 2GB GPU: 128\,s.
\item MSI GF76 Katana with Intel Core i7-12700H 4.7 GHz, 16 GB RAM, Nvidia GeForce RTX 3060 8GB GPU: 41\,s.
\item HP Pavilion 690-00xx with Intel Core i7-8700 4.6 GHz, 16GB RAM, NVIDIA GeForce GTX 1050 2GB GPU: 35\,s.
\end{itemize}

\section{Analysing data in FRELLED 5.0}
\label{sec:anal}
\begin{figure*}[t]
	\begin{center}
		\includegraphics[width=180mm]{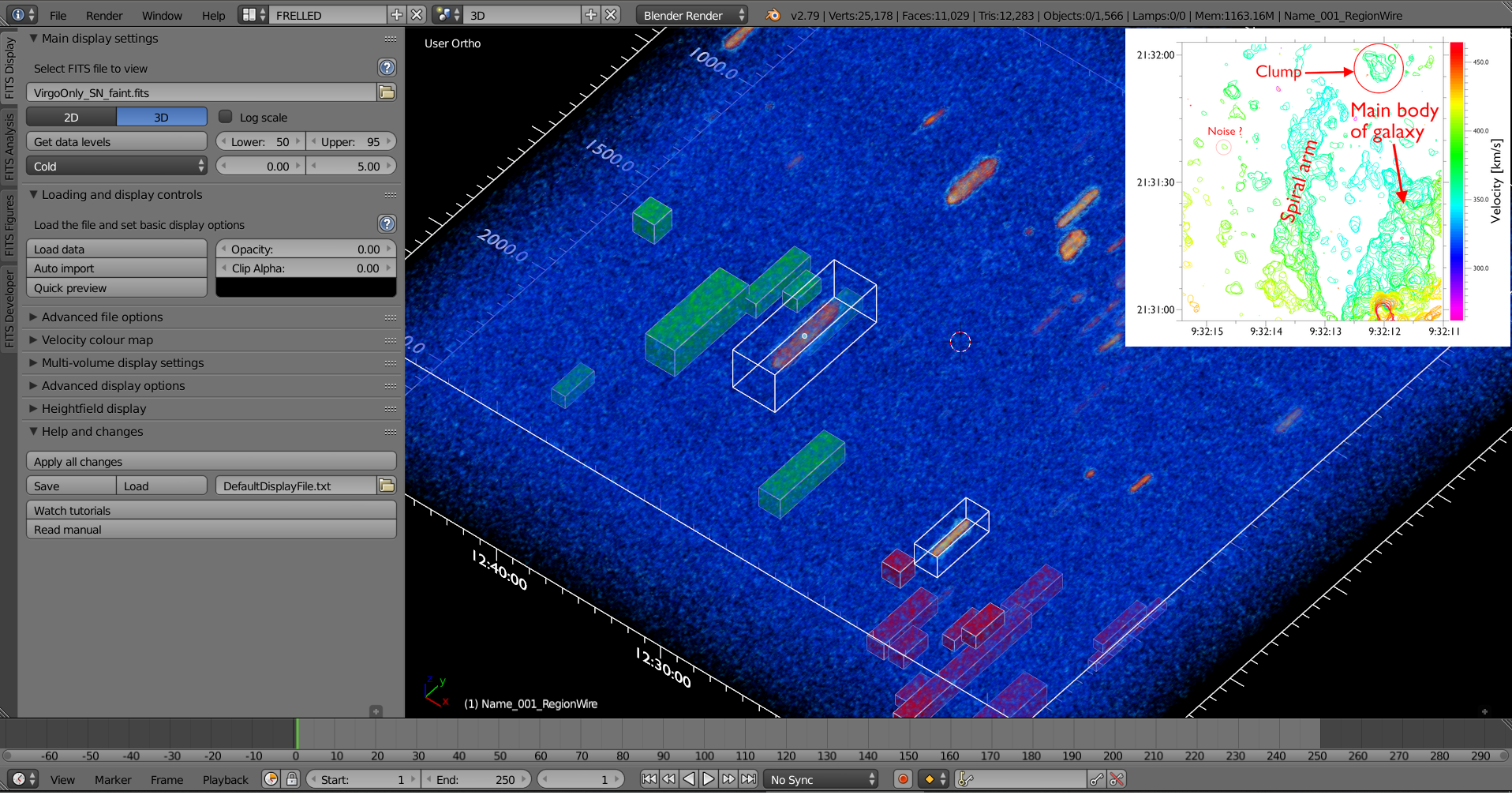}
		\caption[main]{Examples of a variety of FRELLED features. The main GUI is on the left, shown in its default appearance with most menu panels hidden to avoid clutter and to keep things as simple as possible for new users. The main 3D viewport occupies most of the rest of the screen. The file shown is the AGES VC1 data cube described in \cite{GLADoS}. Here, two transfer functions are used, as described in section \ref{sec:multi}, with the faint emission (S/N $<$ 5.0) shown in blue and brighter emission in orange. Some sources have been marked with regions (see section \ref{sec:regions}), here arbitrarily coloured green and red, with two regions shown as transparent wireframes. Finally, the inset image in the upper right demonstrates the capability of creating annotated figures, showing the CO contours from the ALMA PHANGS survey of NGC 2903 (\citealt{phangs}).}
		\label{fig:main}
	\end{center}
\end{figure*}

One of the motivations behind FRELLED is to combine multiple, flexible visualisation techniques together with the capacity for analysis, rather than focusing on one or the other. By `analysis' I here mean the capacity to quantitatively evaluate data and interpret the physics behind it. This section discusses those features most pertinent to that: e.g. the capacity to catalogue data, find the coordinates of visible sources, map their integrated and peak flux, and other features. For the sake of brevity, many of the features described throughout both this and section \ref{sec:interface} are shown collectively in figure \ref{fig:main}.

\subsection{World coordinates}
\label{sec:world}
A major limitation of the previous versions were that many aspects of the world coordinate system (WCS) were hard-coded. In particular, the old code assumed that the rest frequency of the data corresponded to the 21cm emission line of \HI{}. The new system uses both the \textit{astropy} and \textit{spectral-cube} modules. This means that in principle data of any frequency can be loaded, provided the rest frequency is given in the header. If not recognised, FRELLED allows the user to specify the rest frequency manually via a drop-down menu, a text box, and access to the online and exhaustive `Splatalogue' (\citealt{splat}) service that provides a list of lines. Additional deviations from the required keywords may be corrected manually using a supplied external script, see section \ref{sec:scripts}.

As well as \HI{} cubes from a variety of radio telescopes (Arecibo, Westerbork, the Green Bank Telescope, and the Very Large Array), higher-frequency data has also been examined from ALMA (e.g. CO 2-1 from \citealt{phangs}). However since the FITS format does not enforce rigorous standards on the header keywords, it is impossible to guarantee that all data sets will function correctly: in particular, optical interferometry data has not been tested. The default behaviour of FRELLED is to convert frequency to velocity, which is in the (usually) human-friendly values of \kms{}, but this might also not be suitable at high redshifts. Further expanding the capacity for using different WCS will depend on user requests, but the shift away from only support for the \HI{} line is in any case a significant upgrade. FRELLED also still has some hard-coded limits on the range of axes which can be plotted (spatially, all-sky data can be shown, but dimensions less than 2" cause difficulties - axes are shown using sexagesimal notation, and a decimal scheme is not yet implemented), but simply defaults to pixel values if the data are outside these limits. 

Other, smaller changes include a toggle to switch between galactic and equatorial coordinates in the GUI display (though not in the plotted axes), and display both in decimal degrees in addition to the sexigesimal format boxes. Coordinates are also printed to the terminal whenever a navigation query is made, thus giving the user multiple options for copying the coordinates to an external file. The \textit{astropy} module is also used to resolve the sky coordinates of objects specified by name, e.g. entry number in the NGC or other catalogues. Finally, the spectral axis convention can be chosen from the standard optical, radio, and relativistic schemes via a drop-down menu.

\subsection{Regions}
\label{sec:regions}
The core of data analysis in FRELLED remains with its region system. `Regions' are user-created objects which can process selected parts of the data in various ways. The simplest is designating a feature as something of interest, e.g. cataloguing sources visible in the data. Regions can now be assigned different colours, making it easy to visually differentiate by object properties. For example one could specify different colours according to confidence in the signal or object morphology.

Regions can also be used to create various maps. Earlier versions could create integrated flux (moment 0) maps and contours, renzograms, and overlay RGB images from the Sloan Digital Sky Survey (SDSS). The new version also allows peak flux maps and contours, velocity (moment 1) maps, velocity dispersion (moment 2) maps, and isosurfaces (see section \ref{sec:iso}). In addition, renzograms now automatically generate a colour bar indicating the velocity of each contour, minimising the gap between creating a figure useful for quick inspection and something suitable for presentation. 

\subsection{Figures}
\label{sec:figs}
The capacity to generate still figures (and also animations, see section \ref{sec:anim}) has been significantly expanded. In particular, FRELLED now comes with a dedicated menu for creating annotations (text labels, arrows, circles and other symbols) of different colours. Each annotation is associated with a parent region. On rendering a figure, all annotations of the other regions are automatically hidden from view, and this can also be manually toggled. This means that annotations of different regions along the same line of sight do not obstruct each other's view.

Independently of FRELLED development and AGES research, \cite{sports} suggested several features for analysing data cubes that could be useful for visual source extraction. In particular, they noted the value of being able to annotate different features to highlight spurious signals or real but faint objects. With an interdisciplinary approach, they used the `\textsc{SportsCode}' software to act as a `digital notebook' for \HI{} data cubes, allowing expert users to highlight particular features for training new observers. Together with the option to give regions different colours and display them as solid surfaces or with wireframe outlines, most if not all of these features are included in FRELLED. While annotations were not originally intended to be used for training purposes, there is no reason they cannot be used this way. Further discussion of visual inspection training is given briefly in section \ref{sec:sum}.

\section{Interface and documentation}
\label{sec:interface}
Per the design goals set out in section \ref{sec:intro}, the interface of FRELLED is intended to provide astronomer\,\vphantom{0}-\vphantom{}friendly access to Blender. Many of Blender's tools which are required as standard by artists are of no use to astronomers and thus present a potentially confusing distraction. Rather than expecting users to learn the necessary aspects of Blender, the intention with FRELLED has been to group all of the functions necessary for astronomical visualisation and analysis in one location, while features not needed for dealing with scientific data are hidden. In essence, astronomers should be able to learn Blender/FRELLED \textit{as astronomers}, rather than having to become fully-fledged Blender users who use Blender for science.

The change from Blender 2.49 to Blender 2.79 has made this process substantially easier: many features are now accessible via Python which were not previously so; Python-generated GUI panels can be interactively resized, repositioned, and hidden (allowing for customised organisation); from the development perspective, the GUI is generated by the order in which each element appears in the code, rather than the previous incarnation in which the pixel coordinates of each graphical element had to be explicitly specified. All this is aimed at ensuring the interface is easier to learn, use, maintain, and upgrade. The design of FRELLED has also attempted to maximize the data viewing area, giving no more space to its GUI than is required to display the text of the buttons.

In this section I set out those features which were largely possible in previous version of FRELLED, but have had their interface substantially revised and simplified.

\subsection{Display}
\label{sec:dispint}
\subsubsection{Opacity and clipping}
\label{sec:opacity}
While a true interactive colour transfer function is not possible in Blender versions below 2.8, some control of the data display is possible in real time. The maximum opacity level of the data is controlled by a numerical slider : 0.0 making everything transparent and 1.0 making the highest values completely opaque (see section \ref{sec:voldisp} for details). A second slider allows for intensity clipping, where all all values below the specified opacity level are hidden from view without affecting the intensity scaling.

\subsubsection{Multi-volume data}
\label{sec:multi}
Several options in FRELLED allow for the display of multiple data volumes. Chief among these is a dedicated menu, where the user can choose either to generate opacity and transparency from two separate FITS files but loaded as a single volume (multi-component mode), or to render two separate volumes at the same time (true multi-volume mode). The former is anticipated to mainly benefit numerical simulations, e.g. to render opacity from density and colour from temperature (though as noted by \citealt{vohl}, ratios of different spectral line intensities can provide metallicity information, which could also be visualised in this way). The latter is of more use in observational astronomy, allowing different chemical species to be shown at the same time.

Multi-component mode works by generating the opacity and colour maps from the two supplied FITS files, combining them into a single PNG image. Multi-volume mode generates two separate PNG images, one for each FITS file, each with their own opacity and colour maps. These are loaded as separate textures in Blender, with the pre-generated image plane meshes and their materials set to account for this possibility. A GUI menu provides the user control over how these textures are combined to produce the visible image. Importantly, earlier versions of FRELLED showed different PNG images on different plane meshes, significantly increasing the memory overheads and restricting the size of cubes that could be shown in multi-volume mode. The new method overcomes this, allowing multi-volume data to be shown at any size (within the usual limit of the number of image planes, by default 1,500 per projection).
 
There is an additional use for multi-volume rendering. As discussed in section \ref{sec:height}, using standard colour maps it is often difficult to select suitable transfer functions to display data with a high dynamic range. Height maps offer a partial solution to this, but \cite{multicolour} describe an alternative in which multiple transfer functions and colour schemes can be used for different data ranges. The multi-volume capability of FRELLED can be used (with some trivial additional Python code) to provide much the same functionality. The user could, for instance, split the FITS file into two data sets using:
\begin{lstlisting}
image[image < threshold] = math.nan	
\end{lstlisting}	
Where \textit{image} is the 3D array containing the FITS data and threshold is the value below which all data should be replaced with NaN. Then the two data sets can be loaded in the usual way for multi-volume data, with the clipped regions in each file ensuring that their unique colour schemes do not interfere with the display of the other.

The number of data sets that can be displayed volumetrically is limited to two. However, when displaying the data as contours or isosurfaces (section \ref{sec:iso}), no such limit applies. This is possible because FRELLED can process a different FITS file for analysis tasks than the one(s) used for visualisation, as shown in figure \ref{fig:flow}. Generating contours or surfaces require the existence of a region object, and when the user alters their specified parameters (e.g. the data levels to display) and regenerates them, the existing map or surface objects are first removed. However, merely altering the FITS file does not do this: the user must explicitly tell FRELLED to regenerate the map. This means that by creating a series of regions and selecting a different FITS file, it is possible to display an \textit{indefinite} number of different data sets using isosurfaces or contours, a similar capability as is possible in \textit{kvis}.

An example of plotting contours from different FITS files in FRELLED can be seen in \cite{AGESLeo} figure 5. This shows the same original data set both with and without spectral smoothing, i.e. at multiple sensitivity levels and hence with each contour set showing different features. Such processing can be applied to any data set, but variable spatial resolutions and corresponding sensitivity levels are intrinsic to interferometric studies. While it is relatively rare to have multi-wavelength data cubes of the same region, being able to simultaneously display different sensitivity levels presents a much more common use case for multi-volume rendering.

\subsubsection{Velocity maps}
As described above, in multi-component mode the opacity and colours are generated from different data sets. However the `velocity maps' panel in FRELLED allows for a special case of this, in which the colour component is replaced with a colour generated based on position along the spectral axis, i.e. channel number. With this enabled, opacity is still generated from the data values. As in \cite{vohl}, this allows the user to get a better sense of the spectral position of a voxel even in a static, 2D image, without needing to rotate it. Unlike \cite{vohl}, this is limited strictly to the spectral axis, and cannot be recomputed for arbitrary viewing angles.

\subsection{Analysis and figures}
\label{sec:analfigs}
As well as allowing the user to annotate regions for generating figures, a dedicated GUI menu allows users to rescale axes text, to control the background colour, and toggle the display of the background grid. While FRELLED tries to make a sensible initial guess for the appropriate size of the text labels, the ability for the user to alter these makes it much easier to ensure the resulting plot is suitable for their specific purpose. This is especially important when preparing figures for presentation.

\subsubsection{Queries}
\label{sec:queries}
Menus allow queries of the NASA Extragalactic Database (NED) and the SDSS. These can use either the classic or modern NED interfaces, and also restrict objects to galaxies and/or only objects with known redshift measurements. A related feature is that FRELLED can directly overlay SDSS RGB images on the data cube being rendered. For AGES this has been a valuable way to quickly search for optical counterparts and has been an important part in identifying optically dim and dark features. We do not rely on this to determine which objects are definitively unusual at optical wavelengths. Rather, the RGB images (in combination with the query system which can provide optical redshift information) can be used to quickly ascertain if a galaxy has an obvious optical counterpart. Those which do not are then subject to much more detailed examination, i.e. downloading and smoothing the optical FITS files and checking other, deeper surveys where available.

\subsubsection{Miriad}
\label{sec:miriad}
FRELLED can convert regions for use in specific \textsc{miriad} (\citealt{miriad}) tasks. These include the \textit{immask} and \textit{fits} tasks, for masking sections of the data and for converting individual regions between the native \textsc{miriad} and FITS formats, respectively. While FRELLED's regions act as their own masks for data visibility, they do not alter the data itself. The \textit{mask} command lets the user reprocess the data to actually remove the specified regions, which can be useful in statistical analyses of the data properties. Both of these features are implemented as as single-button operations.

A more complex GUI is provided for the \textit{mbspect} spectral line analysis task. By altering the size of the selected region, the user can interactively alter the profile range used for fitting and measuring source properties. For the other parameters, FRELLED provides a GUI to set their values, allowing the user to toggle between the defaults and their own manual values. For example they can set \textit{msbpect} to use different levels of Hanning smoothing, control the order of the polynomial used when fitting the baseline, and select if the calculated flux is a simple summation of the data values accounting for the beam shape, or explicitly assumes the source is unresolved (i.e. \textit{yaxis=sum} is the best choice for extended sources, whereas \textit{yaxis=point} is better for unresolved sources). They can also toggle position-fitting, and FRELLED provides some additional checks beyond \textit{mbspect's} own capabilities. Specifically, \textit{mbspect} will occasionally report a successful fit even when the fitted coordinates are vastly different (several degrees) from the input values. FRELLED includes explicit checks for this, warning the user if the claimed fit is suspicious (more than three pixels away from the input position).

\subsubsection{Animating data values}
\label{sec:anim}
A dedicated GUI panel allows the user to both set up turntable animations (where the virtual camera's view is constantly aimed at some point around which it rotates) and create batch animations of time series of FITS files, i.e. numerical simulations. This panel also offers the capability of animating data display values. Any interpolatable parameters can be animated, for example isosurface contour levels or data display range. The user sets all the initial parameters, enables a button to tell FRELLED that this is the first frame of the animation, and then repeats this for the final parameter values. This generates a series of ASCII display files including an animation toggle parameter, with values for the parameters in the intermediate frames calculated by linear interpolation (or logarithmic interpolation if a logarithmic transfer function is enabled for the data display, and also for the contours). On startup FRELLED always checks for an animation control file and if found and the parameter is enabled, proceeds to automatically load the data with all the specified values, render a figure to a specified location, and then reload itself using the next display file.

While a core design philosophy of FRELLED is that everything should be accessible by GUI, by request it is also possible to generate animations via external script. So long as the animation control file and the display parameter files exist, FRELLED will proceed with the animation. The display files simply consist of the structure:
\begin{lstlisting}
PARAMETER=VALUE 
\end{lstlisting} 
A template script is provided (`External\_AnimationFiles.py') which can be used to set up the display files without ever opening Blender or FRELLED, which includes comments that describe every parameter in detail. 

While within FRELLED the script called from the GUI will only interpolate real numbers, leaving discrete values at their last chosen value, no such limitation exists for the external script. Here one could in principle set a different display colour scheme for every frame if needed. A limitation of the script is that it cannot be run completely independently of FRELLED: at present, FRELLED must still be opened to initially set up a camera. This is a minor restriction though since one could do this for a single initial frame and then specify different FITS files in every frame, thus enabling batch renderings of multiple simulations. In any case, viewing at least one frame of the data manually in FRELLED is always good practise since otherwise it is is very difficult to anticipate the final results.

\subsection{Developer tools}
\label{sec:develop}
A dedicated tab now provides GUI access to tools routinely used in debugging, avoiding the need to access the code itself. This menu also includes several important tools for customising FRELLED for specific use cases, which I briefly describe below.

\subsubsection{Debugging: background scripts}
\label{sec:debugback}
The first panel in the `Developer' menu tab allows control over the scripts which FRELLED executes in the background. The main script in FRELLED controls the visibility of the different image planes according to the orientation of the view, as discussed in section \ref{sec:display}. Scripts which are executed continuously in Blender in this way are implemented via \textit{modal operators}. Toggle buttons here allow ways to both control the active background script and inspect its behaviour. 

First, one toggle button is set according to the value of a boolean variable (in Blender, the link between the variable and the button status is direct, so the button cannot display anything else besides the current variable value). This variable is used throughout FRELLED to effectively (but temporarily) disable any effects of the viewing script, so by inspecting it here, the user or developer can see if this is proceeding in accordance with expectations. Second, another toggle allows the whole modal operator itself to be disabled and re-enabled as necessary, enabling users to forcefully deactivate the operator if they suspect it is interfering with other operations.

The FRELLED menu tabs are all part of Blender's regular tool shelf. By default, the usual Blender tabs (which are primarily concerned with mesh operations) are all hidden. Users can re-enable them in this panel. They can also toggle the view of Blender's header menu, which contains access to many Blender operations (e.g. the layer manager, vertex edit mode settings) which are hidden by default as they are not normally needed.

\subsubsection{Debugging: selection controls}
\label{sec:debugselect}
By default, users cannot select image planes used for displaying the data or other objects critical to FRELLED operations. This prevents users, for example, from accidentally selecting and moving parts of the FITS file around. Similarly the image planes are all hidden by default and only those needed to display the FITS file (one plane per data slice) are shown, to increase performance. In the `selection control' panel, users can override this behaviour, revealing all objects and controlling whether they can be selected or not. Users can also specify and delete individual objects.

\subsubsection{Configuration: deleting objects}
\label{sec:configdel}
As well as controlling whether objects can be selected, this menu also allows users to delete them \textit{en masse}. This is primarily intended for reconfiguring FRELLED for using a different number of image planes, which can be a somewhat lengthy procedure (approximately 30 minutes on the development machine, see section \ref{sec:configvol}). The user has the option to delete the image planes, the user-created objects (e.g. regions and maps), or everything in the file. Since this is an extremely drastic thing to do, multiple toggles must be enabled for this to work, preventing almost any chance of the user doing this accidentally.

These options delete objects and all their associated materials and textures. Special objects present within the file for certain operations (such as the modifier object used for the sum flux procedure, see section \ref{sec:sumflux}) are avoided except in the case of deleting everything. The Python scripts and the file layout are not affected.

\subsubsection{Configuration: volumetric setup} 
\label{sec:configvol}
Once the image planes have been removed as per above, the user can recreate them and their necessary materials and textures, specifying how many are required. By default this cannot be more than 1,000 (which means in practise creating 6,000 objects, see footnote 3 in section \ref{sec:limits}) as Blender's performance scales nonlinearly with the number of objects. On the development machine, creating 500 blanks takes approximately 2 minutes whereas 1,500 takes 30 minutes. An override toggle enables the user to enter larger numbers up to one million.

The default 1,500 pre-generated blank objects was chosen partly based on a balance of performance and data size (see section \ref{sec:sizematters}) but also on familiarity with typical targeted radio observations and smaller surveys. The largest AGES cube, for example, is 620 pixels along its longest spatial axis; THINGS are 1,024 with the largest PHANGS cubes being similar in size. Of course not all data cubes have these dimensions. If necessary, the user can delete the default image planes and create a custom number of objects. The need for higher numbers is obvious in the case of viewing larger files. However, there is also benefit in using smaller numbers of objects. Performance is significantly better in scenes with smaller numbers of objects, so users routinely inspecting smaller data sets may find it worthwhile to remake the master file (see section \ref{sec:filestruc}) using a more suitable number of image planes.

A final option within this section allows users to append objects from an external Blender file. This is designed for the unfortunate case of the user having deleted certain special objects present within the file which are not automatically recreated if destroyed (again especially important for the `sum flux' operation described in section \ref{sec:sumflux}). Here the user only needs to specify the name of the Blender file to append from and the rest is done automatically. FRELLED comes with a `MasterObjects.blend' file for this purpose, so that even if the user has corrupted their main FRELLED.blend file, they can still restore it without needing to download it again. Coupled with the restrictions on deleting objects included in the GUI scripts, this is designed to make the file as robust as possible against user errors while permitting the maximum amount of freedom for users who need to make more drastic changes than are normally required. 
 
\subsection{Documentation}
\label{sec:readthedocs}
Previous versions of FRELLED came with either a PDF manual or an online wiki. This version has expanded this significantly. Perhaps most useful are the help buttons which are present within every GUI panel, providing descriptions of the functionality of each button in that panel. They also describe practical limitations and behaviour which is not immediately obvious, as well as any known issues. This is similarly documented collectively on the wiki (\url{http://frelled.wikidot.com/start}), which also provides video tutorials explaining all the major operations of the software. Both wiki and videos can be accessed directly from within FRELLED itself at the bottom of the main display menu tab. Finally, the much smaller main FRELLED home page (\url{http://www.rhysy.net/Code/Software/FRELLED/}) contains only the essential points regarding installation and basic use.

\subsection{Feature comparisons}
\label{sec:notbugs}
In addition to quantitative performance discussed in section \ref{sec:bench}, it is also useful to compare software based on the different features supported. Having given an overview of the main capabilities of FRELLED in the preceding sections, I here briefly relate these to those of comparable dedicated astronomical tools. I do not consider more general-purpose software such as ParaView, and largely limit the comparisons to the 3D visualisation aspect (one of the primary motivations behind the development of FRELLED). For a recent, much more extensive and thorough overview of astronomical visualisation software, see \cite{survey}.

\subsubsection{GLNemo2}
GLNemo2 is designed for viewing numerical n-body galaxy simulations but can also load spectral-line data cubes. Like FRELLED, these are displayed as a series of image planes using OpenGL shading as Blender does. Another similarity is the motivation for a simple installation procedure (running on Windows, Mac and Linux) and an intuitive GUI, but it differs in being mainly suitable for visualisation rather than analysis of observational data sets. Some powerful visualisation capabilities are included such as the ability to set different parts of the data to display with different transfer functions, and the spectral axis of a data cube can be interactively stretched (this latter is possible in principle in FRELLED but not implemented). Time series of numerical simulations can be viewed volumetrically and the timeline advanced interactively, a feature not possible in FRELLED. Users can toggle between perspective and orthographic modes, as in FRELLED, and set up similar turntable rotations.

GLNemo2 therefore shares many similarities with FRELLED in terms of visualisation, and in some important respects exceeds FRELLED's capabilities (though it lacks displacement maps and isosurfaces). However, its analysis features are mainly aimed towards numerical simulations. World coordinates are not supported and the use of objects is very limited, while moment maps and contours are not available. Conversely, it supports tracking n-body particles and plotting vectors. In short, it is a more powerful tool for analysing numerical simulations, but less so for analysing observational data.

\subsubsection{iDaVIE}
Of the tools considered here, iDaVIE is perhaps the most similar to FRELLED. As already discussed, its quantitative performance in terms of loading and displaying data far exceeds FRELLED. It also allows the creation of moment maps, control over how the spectral axis of a cube is stretched, generation of masks and import of catalogues. The display can also be cropped to specific 3D regions. World coordinates are supported and users can interactively generate regions, and an interesting feature is the ability for users to `paint' voxels to add to mask arrays whose visibility can be toggled on or off. Regions can display statistical information about the data they contain. However, they are not exactly virtual `objects' in the Blender-sense, but are more literally a set of coordinates which define part of the data: they do not have vertices and faces. Generated moment maps can be saved as PNG images, while within iDaVIE itself they are shown in a separate panel rather than in-context as in FRELLED. Contours, channel maps, height maps, isosurfaces, annotations and SDSS maps are not supported.

The most unique feature of iDaVIE is that it operates almost entirely in VR. As well as interacting with the display via controllers (e.g. to translate, rotate, and scale the data), users can also use voice control for many tasks. The interface for interacting with the data in true stereo 3D is highly advanced, but the trade-off is the requirement for a headset. For those with access to VR, iDaVIE is certainly better for viewing large data sets than FRELLED, though its visualisation is limited to volumetric display. The installation procedure is also straightforward but at present it is only available for Windows.

\subsubsection{VisIVO}
The Visualization Interface for the Virtual Observatory comprises a suite of tools, of which the most relevant here is the ViaLactea Visual Analytics tool (\citealt{visivo}). Installation procedure is more complex than the other software considered here as it must be compiled from source along with several dependencies (it is available for Mac and Linux). While aimed at Galactic plane data, with the ability to retrieve online data from the Virtual Observatory and ViaLactea Knowledge Base, it also allows users to examine locally-stored FITS files. It can visualise both 2D data and 3D spectral-line data (as isosurfaces and volumetrically), perform SED fitting as well as plot correlations between chosen variables, and overlay source catalogues from a wide range of available sources at multiple wavelengths. 3D data can be shown as individual 2D slices (though not as height maps) and moment maps can be generated. The session state can be saved, allowing the user to pause their work and return to it later much as with FRELLED. World coordinates are supported. It also provides the facility to merge data cubes of a selected area using Montage (\url{http://montage.ipac.caltech.edu/news.html}).

The VLVA is in some ways more specialised than FRELLED, with available catalogues curated to Galactic centre surveys, while in other ways it offers more general functionality by being equally capable for dealing with 2D and 3D data alike. It does not have the ability to create 3D regions, making it unsuitable for visual cataloguing of spectral line data, whereas FRELLED does not offer the deeper analysis tools such as SED-fitting. This makes the two programs essentially complementary, with the direct overlap in use cases likely small despite some similarities in visualisation capabilities.

\subsubsection{DS9, CARTA, and \textit{kvis}}
Though more dissimilar in terms of visualisation capabilities, three other popular analysis tools are also worth briefly mentioning. DS9 (formally SAOImageDS9, \citealt{ds9}) offers a limited form of volumetric rendering, but each change to the view (translation, zoom, rotation) requires an update to the display that can take several seconds to finalise the render. This, together with the lack of 3D regions, makes it unsuitable for the free-from of 3D exploration and cataloguing for which FRELLED is primarily designed. In other respects, it is far more powerful than FRELLED, allowing multiple FITS files to be shown adjacently or as a blink comparator, as well as being accessible via IRAF and astropy. It is more useful for 2D analysis than 3D examination.

\textit{kvis} (\citealt{kvis}) played the most direct role in influencing the design of FRELLED, especially the ability to rapidly switch  between data projections (a procedure which requires accessing a sub-menu in DS9), the large screen area dedicated to the data display, and the coordinate axes which are designed to allow for easy creation of presentation-quality figures. While extremely fast and allowing for interactive spectra, as well as generating renzograms on the fly, it offers no cataloguing tools, moment maps, or queries to external databases. The \textit{karma} package in which it is included does include a 3D viewer (\citealt{wispy}), but this is only for visualisation, not analysis. 

Finally, the Cube Analysis and Rendering Tool for Astronomy, CARTA (\citealt{carta}) is a powerful visualisation and analysis tool but with a completely different approach to FRELLED. It operates using a client-server structure, allowing a relatively lightweight local machine to display TB-size data cubes. At present it does not offer 3D display. It has a highly customisable GUI designed to allow the setup of multiple windows for comparing data sets, as well as an advanced spectral line analysis tool which allows the user to display multiple spectra (from different or the same data sets) either in adjacent windows or overlaid on one another, and select and measure known lines automatically. Compared to FRELLED, its balance lies strongly in favour of analysis rather than visualisation.

\section{Code and file structure}
\label{sec:code}
All the scripts that comprise FRELLED are Python, with the vast majority being incorporated directly into the main `FRELLED.blend' Blender file. While the user normally sees only the FRELLED GUI (as in figure \ref{fig:main}), it is also possible to access the standard Blender interface and the underlying FRELLED scripts. Python has become enormously popular in the astronomy community, but inevitably many of the FRELLED scripts do require some familiarity with Blender to fully understand. I have tried to provide as much documentation as possible on these for the benefit of those wishing to modify the code.

\subsection{Code structure}
\label{sec:codestruc}
Recoding and debugging FRELLED was performed intermittently throughout 2020-2023. As a solo developer, repeatedly having to rewrite the code like this is simply not practical, and keeping up with the much more frequent alterations to Blender (thanks to its large development team) is just not possible. One manifestation of this is the choice of upgrading from Blender 2.49 to 2.79. At the time recoding began, 2.8 was already released, but its real time display lacked the rapid update of textures in its real time view: any changes to the view resulted in a significant visual lag, hence 2.79 was selected in favour of later versions. 

It is also worth giving an example of how drastic the Python syntax was changed between Blender 2.49 and Blender 2.5+. To create a new object in Blender 2.49, one would use:

\begin{lstlisting}
scene = Blender.Scene.GetCurrent()
ob = scene.objects.new(mesh_structure,object_name)
\end{lstlisting}

Where \textit{mesh\_structure} is the name of an existing mesh structure (a set of vertices and faces) and \textit{object\_name} is a string variable giving the name of the object to be created. "Scene" refers to the Blender scene where the new object will be placed. In Blender 2.5+ the same would be accomplished by:

\begin{lstlisting}
scene = bpy.context.scene
new_object = bpy.data.objects.new(object_name, mesh_object)
scene.objects.link(new_object)
\end{lstlisting}

Virtually all operations, from accessing object location data, to vertex structure and materials, underwent the same level of change. Updating FRELLED to even more recent versions of Blender will be a significant undertaking, however the code has been designed to be much easier to both maintain and migrate. In addition, the Python syntax in later versions of Blender has not undergone the same scale of redevelopment as in the change from 2.49 to 2.79 (with one or two important exceptions, see section \ref{sec:sum}). For example the script written to import the data into Blender 2.91 for virtual reality display (see section \ref{sec:vr}) also functions in Blender 4.0.

The Python scripts in FRELLED 5.0 have a far more modular organisation. Earlier versions used essentially one script per graphical menu, which both constructed the GUI itself and carried out all the operations of each button. The new structure is, as much as possible, to have the graphical scripts limited to constructing graphical elements alone. Each operation then calls a function which in turn invokes a separate script. Some minor exceptions are made for extremely trivial operations (such as altering the value of a single variable) or where Blender's idiosyncrasies necessitate an operation be performed in-place. 

This modular structure has several benefits. Any new developers do not have to wade through thousands of lines of code to find the one operation they may be interested in, making it far easier to modify the code as required. For migrating the code, one in principle need only recode the GUI menu scripts to begin with, and can then proceed in stages to rewrite each operation. This would make it much faster and easier to get at least some level of functionality, whereas the previous version of FRELLED demanded an all-or-nothing approach. I note also that the basic syntax of generating GUI menus in Blender has not changed between versions 2.79 and later ones.

Finally, a standalone script (not called by any others or accessible with the GUI) `FindOriginOfScripts.py' finds where each script is called from. This information is listed at the top of each script as a comment, describing whether the script is imported as a function or executed as a script and where in the code this is done. Similarly another script `FindStringAllScripts.py' allows the user to search all scripts in the FRELLED.blend file for a given string. In addition, an "Essential Notes" file within the FRELLED Blender file describes several routine operations needed throughout the code for ease of copying frequently-used snippets, as well as methods for executing various functions that have been tried and rejected. All scripts are thoroughly commented.

\subsection{File structure and installation}
\label{sec:filestruc}
FRELLED is implemented as a downloadable Blender file in a zip archive containing several other files: Python scripts, default values in ASCII files, an example FITS file, and the text files for the individual panel help buttons. It also contains two json files required for \textit{matplotlib} which are not always present or created correctly (this is documented on the wiki).

The installation procedure for FRELLED has been tested on multiple machines running both Windows (version 10 and 11) and Linux (Debian and Ubuntu, but this should not matter). It has been designed to be much simpler than the previous version, an important point since in some cases this was extraordinarily difficult. The new procedure, in essence, is a short, three-step process:
\begin{enumerate}
\item Download and unpack a special version of Blender which includes the additional Python modules needed for FRELLED.
\item Download and unpack the FRELLED files.
\item Run a simple, one-line command to configure Blender, and set up an alias or shortcut to run it.
\end{enumerate}
The most important improvement in these steps is the first. Previously, users would have to manually install all the Python modules needed into Blender's supplied standalone version of Python. On certain Linux networks where the system Python could not be circumvented this proved ultimately impossible, and Python managers such as Anaconda are ill-equipped to deal with the Blender environment. None of this is now necessary. By installing the modules needed into local installations of Blender's Python, the entire Blender installation can be compressed and distributed as an archive. Since the installation process of the end user just consists of unpacking this archive, no interference with their own existing Python installation occurs, and they do not have to install any Python modules themselves as these are all provided in the archive. The other two steps are trivial.

Once installed, as previously the FRELLED directory can be considered a master copy which users are advised not to use directly. Instead they should copy this entire directory and place the FITS files they wish to view in the copy. This ensures the original files are always present and correct, but it also prevents possible naming conflicts. Some operations in FRELLED create supplementary files based on the region names, and as well as this becoming potentially confusing, it also carries the risk of files being overwritten.

An alternative would have been to make FRELLED a more standard `add-on' for Blender. The advantage of this, in principle, would be that users could toggle in effect a `FRELLED mode' from within any Blender file and then create project directories anywhere they wished, as well as avoiding the need for step 3 above. This was rejected because of the great advantage of providing a Blender file containing the pre-generated blank image planes. Since it is extremely slow even to import these into a new Blender file (see section \ref{sec:configvol}), implementing FRELLED as an add-on would give it impractical start-up times. 

Another option would be to have all of FRELLED's Python scripts stored as external files in some master directory, with each instance of FRELLED using this central repository. The advantage of this would be that one could edit scripts across multiple FRELLED files simultaneously. The disadvantage would mean that backwards compatibility would become absolutely unavoidable, even when changing the behaviour of a feature would be desirable, and bugs would then apply to all a user's files. The chosen system of keeping FRELLED in self-contained files has the advantage of making it easy to have multiple versions running alongside one another, and development can proceed without enforcing backwards compatibility. 

A further benefit of this structure is that users only need to complete steps one and three whenever the required Python modules are changed, which is anticipated to be highly infrequent. Once users have done this, upgrades to FRELLED will almost always consist simply of step two: downloading the new Blender files. 

\subsection{A new method for summing the flux}
\label{sec:sumflux}
One change which is invisible to the user is the operation to sum the flux within arbitrary-geometry volumes. Previous versions of Blender had built-in tasks in Python to accurately assess whether a point at any given coordinates was inside a closed mesh or not. Extensive testing of multiple techniques purported to do the same in Blender 2.79 has shown that the accuracy of this is significantly worse than in the earlier versions. Therefore a new method was devised.

In regular, non-rotated cuboids, summing the total flux (or more generally, data values) present is trivial: an array can be extracted from the data given the specified coordinates directly using \textit{numpy} and the \textit{nansum} function. Where the mesh geometry is arbitrarily complex, one must first define every voxel which is inside the volume and sum only those values. This is done by means of a special object (hidden from the user) within the file that has a `remesh' modifier applied. This is a built-in Blender tool that converts any mesh into cubical blocks. The scaling and resolution of this modifier have been chosen to ensure that the blocks are of integer pixel sizes. This is then copied and applied to the target mesh the user wishes to use for summing the flux.

\begin{figure}[t]
	\begin{center}
		\includegraphics[width=90mm]{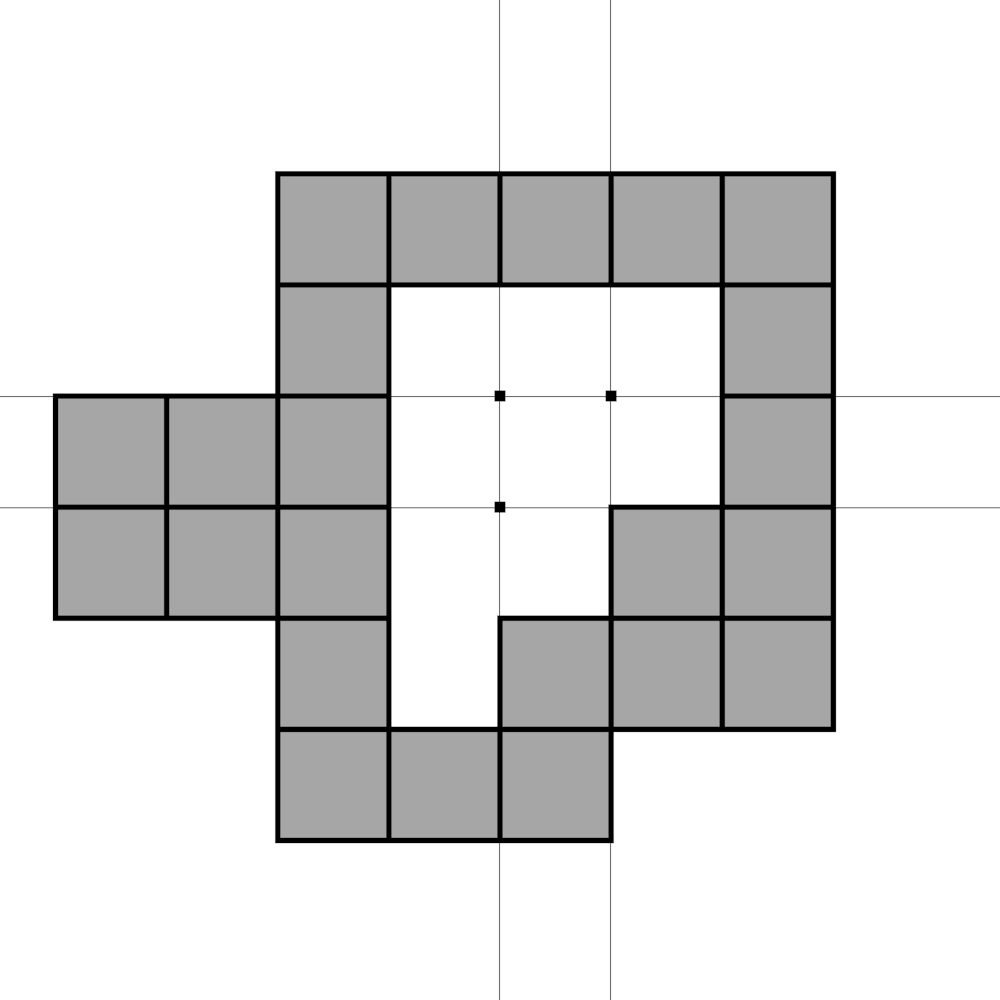}
		\caption[sum]{Illustration of how FRELLED computes the total flux within arbitrary volumes, here showing a two-dimensional slice for simplicity. First, a Blender modifier is used to convert the mesh into a cubical geometry, shown by the filled grey regions with the polygons as thick black lines. The problem is then analysed one-dimensionally. Proceeding along any of the thin black lines (horizontally or vertically makes no difference), the algorithm checks for the presence of a vertex at the current and next position. Once a gap is found after a pair of adjacent vertices, it marks the location to fill in, stopping once a new adjacent pair is encountered. The black squares show the new vertices that would be created by this procedure.}
		\label{fig:sumflux}
	\end{center}
\end{figure}

The Blender modifier applies only to the surface of the mesh. The next step is to fill in the interior. This is done by searching for alternate pairs of vertices. In a closed mesh, going along any axis, the first pair of vertices encountered must define the cube on the surface of the mesh, and, after a gap, the next pair must therefore define the opposite surface. Additional vertices are then generated at all integer points between these pairs which are not already occupied (see figure \ref{fig:sumflux}). This process is imperfect but performs better than the alternative methods examined, generally only returning small (few pixel) errors even on complex meshes such as the built-in Blender monkey primitive. On simpler meshes errors are rare. The resulting vertex cloud that this generates is visible and editable by the user, so those more familiar with Blender can make manual adjustments if high accuracy is required.

This method, like all operations which require direct access to the FITS data (e.g. mapping, see section \ref{sec:regions}), is robust to sparse sampling. Regardless of how many image planes are displayed, they are placed at the same coordinates, that is where one pixel (or spectral channel) is equivalent to one Blender Unit.

\subsection{Additional scripts}
\label{sec:scripts}
FRELLED includes a few external Python scripts suitable for processing the FITS files prior to visualisation. The `Normalise.py' script scales the data according to the minimum and maximum values in each spectral channel, designed for data sets with emission of varying dynamic range across the whole volume, e.g. Galactic \HI{}. The `FixAxes.py' script simply reshapes 4D data to 3D.  

The `AGESCleaning.py' script, originally developed for AGES data cubes, is more versatile. This can fit and subtract \textit{n}\textsuperscript{th}-order polynomials along the spectral axis (this is done for each spectrum i.e. along each spatial pixel), optionally rescaling them by the spectra's \textit{rms}. Normalising by \textit{rms} is a common technique for source finding (e.g. \citealt{SOFIA}, \citealt{saint}) but also useful in visualisation (\citealt{agesvii}, T15). The script can also reprocess the spatial bandpass using the MINMED and MEDMED methods described in \cite{minmed} and \cite{agesvii}; see also \cite{isolated}. In brief, the spatial bandpass is subdivided into (typically) five sections and the \textit{rms} measured within each, with the value of either the minimum of the medians or the median of the medians then subtracted from the whole bandpass. 

Finally, the script can also edit header keywords. At various stages, FRELLED includes protections warning the user if required keywords are not found, directing them to this script if this cannot be avoided. One example is the spectral axis. The \textit{spectral-cube} module requires this to be exactly `Hz'\footnote{The versions of the modules FRELLED uses have been carefully chosen to ensure they are all compatible with each other, under the constraint that some later versions are not accessible to Blender's version of Python. It is likely that later versions are more robust to some situations than the versions used by FRELLED.} , and it is simpler to just ask the user to edit this if necessary rather than anticipate every possible variation that could be employed. As mentioned in section \ref{sec:world}, the FITS format does not impose rigorous standards, so having the facility for the user to make the necessary edits themselves is the safest and most robust fallback option.

\section{Summary and future work}
\label{sec:sum}
FRELLED version 5.0 has been redesigned from the ground up. At all levels, it should be simpler to both install and use. The installation procedure in particular is now reduced to (essentially) downloading two files and running a single command from a terminal. Zero prior experience of using Blender is assumed. Where necessary, this has meant replicating functionality of standard Blender buttons in order that they are conveniently accessible from the FRELLED GUI menus - users need never see the standard Blender interface at all. Options normally present in the tool shelf FRELLED uses for its GUI have been disabled so that the focus is exclusively on astronomical data visualisation. Tasks which were previously possible only by carefully reading the documentation, such as multi-volume rendering, are now explicitly included as GUI menus. Tasks which could only be done by prior familiarity with Blender, such as making renzogram colour bars, are now fully automated.

This simplification of interface (or adoption of `fastronomy') is essential. The interface changes in FRELLED are designed so that users will \textit{actually use} the available features, rather than merely making it possible in principle.

FRELLED 5.0's new visualisation capabilities are isosurface rendering, height maps, and virtual reality display (via export). In addition, the speed of viewing both volumetric and 2D data has been greatly increased (typically by a factor of a few). Hard-coded aspects restricting the WCS to \HI{} data have been generalised to allow data of all frequencies and allow conversion between equatorial and galactic spatial coordinates. New maps can be generated to analyse regions of specific interest, such as velocity dispersion and peak flux maps, figures can be easily annotated, and data values can be animated via the GUI or via script. Changes in the interface provide an interactive adjustment of the opacity, a dedicated multi-volume GUI which can also be used to render a single data set using multiple transfer functions, and all aspects of the \textit{mbspect} spectral analysis tasks are now accessible in the GUI. Users and developers now have GUI access to tools routinely used for configuring and debugging FRELLED, without needing to inspect the code.

Changes to the interface may have additional, unexpected benefits. In earlier versions of FRELLED it was possible to annotate parts of the data cube only with knowledge of Blender's artistic tools, whereas FRELLED 5.0 provides a dedicated, astronomer-friendly GUI to make this easier. The intention was to simplify the creation of figures for presentations, but \cite{sports} described the use of annotations for training new observers in visual inspection. In AGES we have not thought to use annotations in this, but instead train new observers by using regions which mark known real and/or spurious detections, providing essentially the same functionality by a different method.

In comparison with other software, FRELLED is intended to provide a balance between visualisation and analysis capabilities. Unusually, it provides tools to visualise data using multiple techniques, and its regions are designed to be easily created and manipulated in 3D. In terms of analysis it does more than some but less than others, allowing (for example) spectral line analysis but not SED-fitting, queries to external databases but not automatic catalogue overlays from online databases. The modular structure of the code, however, makes adding additional analysis capabilities relatively straightforward: in principle, any operation possible in Python can be applied to selected regions and a GUI-wrapper created to support it.

In terms of future development, the modular structure of the code is intended to greatly flatten the learning curve. Eventual migration to Blender version 4+ would not require the same level of almost-complete recoding necessary between Blender versions 2.49 and 2.5+. The chief changes relevant to FRELLED are that Blender 4+ uses a more powerful \textit{collection} system rather than \textit{layers} for managing the display of objects, and a different `EEVEE' real time display engine requiring different materials (though scripts in FRELLED already allow conversion between the material systems). There are doubtless others, such as the selection state of objects now being a function rather than a property, however none of these are sufficient to require a full recode. 

Migrating FRELLED to the latest Blender version would bring several immediate benefits. The EEVEE materials, unlike the old Blender internal display engine, can be fully updated in real time, allowing the colour transfer function to be changed without needing to reload the data. They also do not require different objects for opposing viewing angles, reducing the number of required objects in the scene by a factor two (allowing for improved performance and/or an increase in the permitted size of the data sets). Preliminary testing suggests the increased performance may even be enough to simply load all three sets of image planes into one collection, avoiding the need for any background script to update the images shown depending on viewing angle. Additionally, virtual reality can be enabled directly without needing any additional operations (plugins are available to allow object editing in Blender's VR display, potentially enabling full analysis in VR within FRELLED). Multiple colours for particle systems are also possible, should there be user interest in restoring this feature for viewing n-body simulations. On the longer term, keeping FRELLED within the latest version of Blender is essential for bug fixing and access to new and improved data visualisation capabilities.

\section*{Data availability}
All data sets shown in this work are publicly available online: the PHANGS data can be accessed at \url{https://www.canfar.net/storage/list/phangs/RELEASES/PHANGS-ALMA/}, the THINGS data sets are available at \url{https://www2.mpia-hd.mpg.de/THINGS/Data.html}, and the AGES M33 cube is included in the FRELLED download package. Due to the closure of Arecibo Observatory the original AGES data archive is currently offline but requests for data access can be sent to the author. All the code of FRELLED is included in the download and accessible within the main `FRELLED.blend' file and a series of Python files. Blender itself is open source and the code can be downloaded from the Blender website, \url{www.blender.org}.

\section*{Acknowledgements} 

I wish to thank the numerous volunteers who helped beta test the code and especially the installation procedure, which was often a less than pleasant experience for all concerned. Special thanks to Robert Minchin, Emily Moravec, Romana Grossová, and Vojtěch Part\'{i}k. I thank also the two anonymous reviewers, whose detailed comments substantially improved the manuscript.

This work was supported by the institutional project RVO:67985815 and by the Czech Ministry of Education, Youth and Sports from the large lnfrastructures for Research, Experimental Development and Innovations project LM 2015067. 

This research has made use of the NASA/IPAC Extragalactic Database (NED) which is operated by the Jet Propulsion Laboratory, California Institute of Technology, under contract with the National Aeronautics and Space Administration.

This work has made use of the SDSS. Funding for the SDSS and SDSS-II has been provided by the Alfred P. Sloan Foundation, the Participating Institutions, the National Science Foundation, the U.S. Department of Energy, the National Aeronautics and Space Administration, the Japanese Monbukagakusho, the Max Planck Society, and the Higher Education Funding Council for England. The SDSS Web Site is http://www.sdss.org/.

The SDSS is managed by the Astrophysical Research Consortium for the Participating Institutions. The Participating Institutions are the American Museum of Natural History, Astrophysical Institute Potsdam, University of Basel, University of Cambridge, Case Western Reserve University, University of Chicago, Drexel University, Fermilab, the Institute for Advanced Study, the Japan Participation Group, Johns Hopkins University, the Joint Institute for Nuclear Astrophysics, the Kavli Institute for Particle Astrophysics and Cosmology, the Korean Scientist Group, the Chinese Academy of Sciences (LAMOST), Los Alamos National Laboratory, the Max-Planck-Institute for Astronomy (MPIA), the Max-Planck-Institute for Astrophysics (MPA), New Mexico State University, Ohio State University, University of Pittsburgh, University of Portsmouth, Princeton University, the United States Naval Observatory, and the University of Washington.

\bibliographystyle{aa}
\bibliography{references}

\end{document}